\documentclass{article}

\usepackage{PRIMEarxiv}

\usepackage[utf8]{inputenc} % allow utf-8 input
\usepackage[T1]{fontenc}    % use 8-bit T1 fonts
\usepackage{hyperref}       % hyperlinks
\usepackage{url}            % simple URL typesetting
\usepackage{booktabs}       % professional-quality tables
\usepackage{amsfonts}       % blackboard math symbols
\usepackage{nicefrac}       % compact symbols for 1/2, etc.
\usepackage{microtype}      % microtypography
\usepackage{lipsum}
\usepackage{fancyhdr}       % header
\usepackage{graphicx}       % graphics
\graphicspath{{media/}}     % organize your images and other figures under media/ folder
\usepackage{amsmath}
\usepackage{amsfonts}
\usepackage{graphicx}
\usepackage{parskip}
\usepackage{comment}
\usepackage{csquotes}
\usepackage{geometry}
\usepackage{array}
\usepackage{booktabs}
\usepackage{subcaption}
\usepackage{tabularx}

\usepackage{xcolor}
\definecolor{ForestGreen}{RGB}{34,139,34}
\definecolor{MidnightBlue}{RGB}{25, 25, 112}
\hypersetup{
    colorlinks=true,
    urlcolor=MidnightBlue,
    linkcolor=ForestGreen,
    citecolor=ForestGreen
}

\usepackage[
backend=biber,
style=authoryear,
]{biblatex}

\newcommand\blfootnote[1]{%
  \begingroup
  \renewcommand\thefootnote{}\footnote{#1}%
  \addtocounter{footnote}{-1}%
  \endgroup
}

\def\sym#1{\ifmmode^{#1}\else\(^{#1}\)\fi}

\title{Estimating Idea Production: A Methodological Survey}
\author{
  Ege Erdil \\
  Epoch AI\\
  \texttt{ege@epochai.org} \\
   \And
  Tamay Besiroglu \\
  Epoch AI\\
  \texttt{tamay@epochai.org} \\
  \And
  Anson Ho \\
  Epoch AI\\
  \texttt{anson@epochai.org}
}

\begin{document}
\maketitle

\begin{abstract} 
Accurately modeling the production of new ideas is crucial for innovation theory and endogenous growth models. This paper provides a comprehensive methodological survey of strategies for estimating idea production functions. We explore various methods, including naive approaches, linear regression, maximum likelihood estimation, and Bayesian inference, each suited to different data availability settings. Through case studies ranging from total factor productivity to software R\&D, we show how to apply these methodologies in practice. Our synthesis provides researchers with guidance on strategies for characterizing idea production functions and highlights obstacles that must be addressed through further empirical validation.\blfootnote{We thank Chad Jones, David Roodman, Carl Shulman, Mike Webb and Jaime Sevilla for their helpful comments. You can find the code for experiments in \href{https://github.com/ege-erdil/estimating-returns-to-rnd}{this GitHub repository}.}
\end{abstract}

\section{Introduction}

A core insight of modern growth economics is that ideas are crucial for developing new technologies that drive economic growth \parencite{romer1990endogenous, jones1995r}. Modeling the production of ideas is thus crucial, both for determining the optimal allocation for innovation activity \parencite{arrow1972economic}, and for determining the dynamics of an economy's growth. 

Despite its importance, estimating idea production functions in practice is riddled with difficulties. For example, there may be insufficient high-quality data on inputs and outputs, model misidentification, and strong correlations between input and output measures. These issues can make resulting estimates highly uncertain or unreliable.

In this paper, we take a step towards addressing these issues. In particular, we provide a comprehensive methodological survey of strategies for estimating idea production functions, summarized in Table \ref{tab:estimation-approach-summary}. We focus on estimating the law of motion for total factor productivity (TFP) introduced by \cite{jones1995r} (hereafter referred to as \textit{the Jones law of motion}), versions of which have been used to measure the returns to R\&D in multiple domains (e.g. \cite{bloom2020ideas}). Each strategy is supported by mathematical derivations to either validate the estimation approach, or clarify the assumptions necessary for the method to be applicable. 

% While the law appears simple conceptually, its practical use tends to be anything but. Estimation challenges include misidentification of the model, lack of abundant high-quality data on inputs and outputs, and potential correlations between technology $A$ and $I$.

\begin{table}[h]
\centering
\small
\begin{tabularx}{\textwidth}{ X X X X }
\toprule
Method & Summary & Accuracy & Reference \\
\midrule
Naive &  { \large $\frac{\text{Output growth rate}}{\text{Input growth rate}}$} & Rough estimates & \cite{bloom2020ideas} \\
\addlinespace
Linear Regression & Log-linear approximation & Good with sufficient data & \cite{Pessoa2005IdeasDG} \\
\addlinespace
Max Likelihood Estimation & Stochastic model + MLE & Best with abundant data & \emph{Our contribution} \\
\addlinespace
Bayesian & Stochastic model + Priors & Useful with limited data & \emph{Our contribution} \\
\bottomrule
\end{tabularx}
\caption{Summary of our methods for estimating the returns to R\&D.}
\label{tab:estimation-approach-summary}
\end{table} 

Statistical techniques, while valuable, are insufficient to fully address the fundamental challenges of measurement and model uncertainty. To assess the practical applicability of these theoretical estimation approaches, we conduct comprehensive empirical case studies in three domains: US total factor productivity (TFP), computer chess, and various other areas of software R\&D. These analyses shed light on the obstacles commonly encountered in real-world settings, serving as both warning signs of potential pitfalls and compelling evidence of the need for enhancements to existing empirical practices.

\subsection{Prior work} 
\label{sec:prior-work}

There have been many works attempting to estimate key parameters of the idea production function since its introduction \parencite{Hall2009MeasuringTR}. \textcite{Neves2018SpilloversIT} perform a meta-analysis of several independent estimates, finding weakly diminishing returns to having a larger stock of ``ideas" over time. \textcite{Sequeira2020SteppingOT} perform a similar meta-analysis for R\&D inputs, finding fairly strong diminishing returns to scale from increasing inputs -- this can be interpreted as a ``stepping on toes" effect where research effort cannot easily be parallelized. 

An important decision in any paper estimating the returns to R\&D is data selection -- which metrics are chosen to measure the quantities of inputs and outputs of the production function. Typical examples of input measures include the dollars devoted to R\&D \parencite{Furman2000TheDO, Bloom2005IdentifyingTS} and the number of full-time equivalent researchers \parencite{Porter2000MeasuringT, Pessoa2005IdeasDG, Luintel2009HeterogeneousIP}. Common output measures are the number of patents \parencite{Porter2000MeasuringT, Lanjouw2004PatentQA} and TFP \parencite{bloom2020ideas, Herzer2020SemiendogenousVS}. As we discuss in Section \ref{sec:measurement-challenges}, there are many difficulties that arise when trying to choose these measures in practice. 

Besides choice of data, another crucial degree of freedom lies in the estimation approach. For example, \textcite{bloom2020ideas} estimate research returns by taking the ratio of the growth rate in outputs $g_A$ to the growth rate in inputs $g_I$. On the other hand, \textcite{Pessoa2005IdeasDG} considers a log linear approximation to the law of motion from \cite{jones1995r} and estimate its parameters using linear regression. As we will see in Section \ref{sec:estimation-strategies}, these methods are either limited in applicability or accuracy. A crucial motivation for this paper is thus to present two alternative estimation approaches based on stochastic models, which can help circumvent these issues.  

\section{Background}
\label{sec:background}

\subsection{The Jones law of motion}
\label{sec:jones-law}

The study of returns to research effort starts with the law of motion introduced in \textcite{jones1995r}.\footnotemark While it was originally used to model TFP \( A \), it has since been used to model improvements for specific technologies, such as computing hardware, agricultural, and medical innovations. This law of motion is specified by
\begin{equation}
    \frac{1}{A} \frac{dA}{dt} = 
    % \theta A^{\phi-1} I^{\lambda} = 
    \theta A^{-\beta} I^{\lambda},
\end{equation}
where \( I \) is some measure of inputs to R\&D and \( \theta, \beta\), and \(\lambda \) are model parameters. The model captures two important effects: increasing versus diminishing returns on finding new ideas over time, quantified by \( \beta \), and the returns to scale on research effort at any given instant, quantified by \( \lambda \).
\footnotetext{The Jones law of motion generalizes many that have previously been used in the endogenous growth literature. Steady-state exogenous growth models correspond to \( \beta = \lambda = 0 \), while the classical endogenous growth model from \cite{romer1990endogenous} corresponds to \( \beta = 0, \, \lambda = 1 \). \cite{bloom2020ideas} describe how their evidence on ideas getting harder to find is consistent with a model having \( \beta, \lambda > 0 \).}

The interplay between these two effects is best described by the parameter \( r = \lambda/\beta \), which \cite{bloom2020ideas} calls the \textit{returns to research effort}. This plays a central role in determining the asymptotic properties of any endogenous growth model that features this law of motion.\footnote{We provide more detail on the importance of $r$ in endogenous growth models in Appendix \ref{sec:importance-of-r}.} As we describe in Section \ref{sec:naive-method}, this is because \( r \) can be equivalently characterized as the ratio between the growth rate of \( A \) and the growth rate of \( I \) \emph{in a steady-state growth equilibrium} in which both quantities grow exponentially. 

The law is appropriate to use whenever we think an exponentially growing trajectory in the inputs \( I \) should eventually lead to exponential growth in productivity \( A \), possibly after an initial period of convergence to equilibrium. This principle can be used to decide which parametrization for \( A \) makes the most sense, as it might not be clear for some time series \( X \) having to do with efficiency in some domain whether \( X, \log X, \text{ or } \exp(X) \) is a better choice for \( A \).

\subsection{Preliminaries}

While the Jones law of motion is conceptually simple, obtaining accurate estimates of $r$ can be quite complicated. The methods that we choose in practice will depend on the particular domain under study. In this subsection we briefly introduce the main mathematical approaches that we use in the context of a real-world case study -- empirical estimates can be found in Section \ref{sec:case-studies}.

\begin{figure}[h]
    \centering
    \includegraphics[scale=0.75]{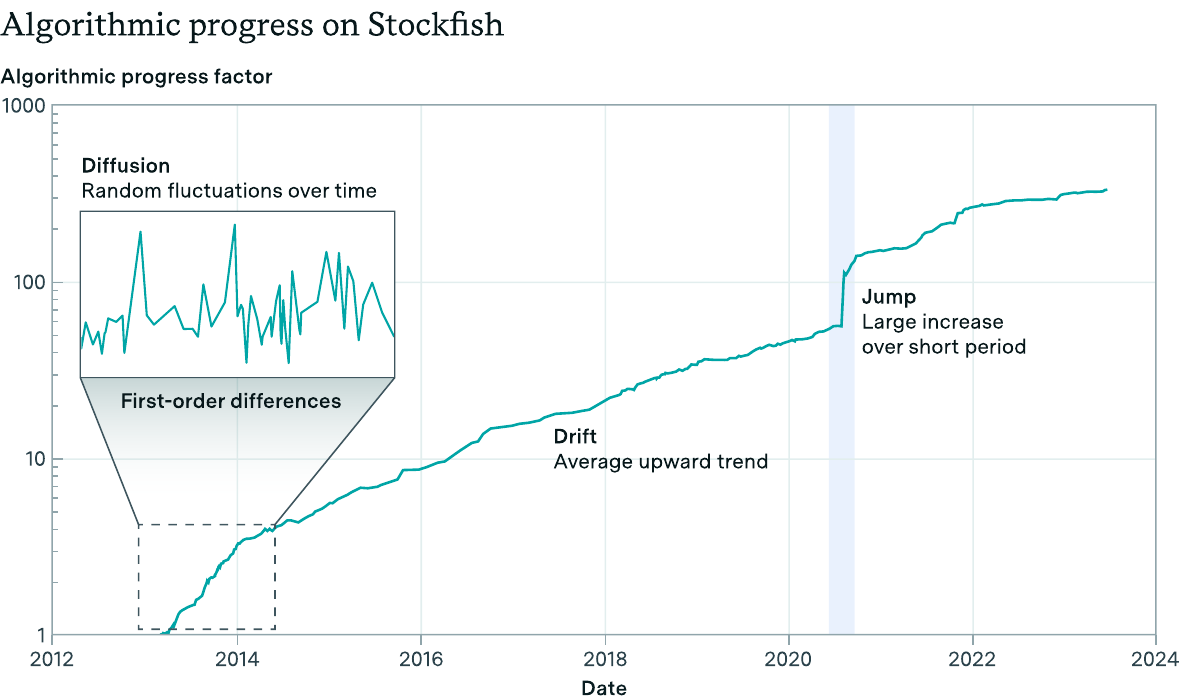}
  \caption{Progress in the algorithmic efficiency of Stockfish over time.}
  \label{fig:stochastic-process}
\end{figure}
\subsubsection{Software efficiency}

Between 2013 and 2024, the Stockfish open-source chess engine has been frequently updated to increase performance (e.g. as measured in Elo score). As new improvements are introduced over time, Stockfish can achieve the same level of performance with fewer computational resources (or less running time), resulting in an improvement in ``\emph{software efficiency}".\footnote{Note that in general, software efficiency improvements may occur in complementary fashion with improvements in hardware \parencite{Hooker2020TheHL}.} We illustrate these algorithmic improvements in Figure \ref{fig:stochastic-process}, where the overall gain is the total reduction in computational resources (or runtime) required to achieve the same level of performance compared to previous dates.

\subsubsection{Stochastic calculus}
The time series shown in Figure \ref{fig:stochastic-process} illustrates small local fluctuations (``diffusion") but with an upward ``drift" over time. In addition to these two properties, we also observe a salient ``jump" in the algorithmic progress factor in 2020.\footnote{This was due to the introduction of neural-network based methods into Stockfish.} These constitute three key properties of the time series that we want our models to account for. 

To capture the first two of these three properties, we can model the time series as a ``\emph{drift-diffusion process}" $X_t$. One example of this is the \emph{Wiener process} $W_t$ with drift, which has been used to model the particle Brownian motion (among other phenomena). This can be described by a stochastic differential equation with a corresponding diffusion and drift terms:
\begin{equation}
    dX_t = \underbrace{\mu \: dt}_{\text{drift}} + \underbrace{\sigma \: dW_t}_{\text{diffusion}},
    \label{eq:brownian-motion}
\end{equation}
where $\mu$ is the mean of $X_t$, and $\sigma$ is the standard deviation.\footnote{In some models these are simply constants, but in general $\mu$ and $\sigma$ may depend on $X_t$ and $t$.} In some cases (e.g. analyzing stock prices), a more suitable model would be to model the \emph{logarithm} of some time series as a Wiener process, also known as ``geometric brownian motion". In this case, we replace $dX_t$ with $dX_t / X_t$ in equation \ref{eq:brownian-motion}.

More generally, we may wish to express the stochastic differential equation in terms of a function of $X_t$. Concretely, consider a function which takes values $f(t,x)$ for real $t$ and $x$. To obtain an expression for $df(t, X_t)$, we need to use a modified version of the chain rule that works with stochastic processes, known as \emph{Ito's lemma}. In general this is derived using a Taylor expansion and noting that the Wiener process has quadratic variation $dW_t^2 = dt$. For the purposes of this paper the specific form of the lemma we use is given by 
\begin{equation}
    df = \left( \frac{\partial f}{\partial t} + \frac{1}{2}\frac{\partial^2 f}{\partial x^2} \right) dt + \frac{\partial f}{\partial x} dW_t.
\end{equation}

One downside of the Wiener process is that it does not capture the third property in Figure \ref{fig:stochastic-process}, i.e. ``jumps". To model this, we generalize from Wiener processes to \emph{Lévy processes}. Each Lévy process can be thought of as a random process with both a drift-diffusion component and a jump component, allowing us to model all three of the aforementioned properties. 

\subsubsection{Special cases}
\label{sec:special-cases}

One particular set of Lévy processes that will be useful to consider is the family of \textbf{\emph{stable} Lévy processes}. Rather than considering purely special cases where the increments of the random process are normally distributed, we consider the more general family of \emph{stable distributions}. In particular, if a linear combination of independent random variables $X_i$ following some distribution itself follows the same distribution (up to some scale and shift), then we say the distribution is stable. 

They are characterized by four key parameters: stability $\alpha$, skewness $\beta$, rate $\mu$, and scale $c$. These control the heaviness of tails, degree of asymmetry, location of the median, and spread of the distribution respectively. If a random variable $X$ is stable distributed, we write $X \sim \text{Stable}\left(\alpha, \beta, \mu, c\right)$. 

For our purposes, these distributions are useful to consider because they help model empirical observations that are more heavy-tailed than the Normal distribution, and also are theoretically supported by certain generalizations of the Central Limit Theorem \parencite{borak2005stable}.

Another useful special case is when $dX_t/X_t$ is \textbf{\emph{scale invariant}}, such that the dynamics of the process do not depend on specific time scales. An example of a stochastic differential equation that has this property is the Cox–Ingersoll–Ross (CIR) model of interest rates from \cite{cox2005theory}, given by
\begin{equation} 
\frac{dr_t}{r_t} = -a \, dt + \frac{ab}{r_t} \, dt + \frac{\sigma}{\sqrt{r_t}} \, dW_t 
\end{equation}
with \( r_t \) denoting the rate of interest at time \( t \), \( a, b, \sigma \) are constants, and \( dW_t \) is a Wiener process. The important detail for our purposes is that due to the quadratic variation $dW_t^2 = dt$, the noise term is multiplied by \( \sqrt{1/r_t} \) instead of some other factor. This is because this particular specification makes the action of \( r_t \) analogous to changing the scale of time: if we dilate a drift-diffusion process by a factor \( f \), the drift term is multiplied by \( f \) while the diffusion term is multiplied by \( \sqrt{f} \).

One way in which the CIR model can be solved is by letting $r_t$ be a \emph{Feller diffusion} process. In general, $X$ is a Feller diffusion if it has a stochastic differential equation given by
\begin{equation}
    dX = c dt + \sigma \sqrt{X} dW_t,
\end{equation}
where $c$ is a drift factor, $\sigma$ is a constant, $t$ is time and $W_t$ is a Wiener process. The important feature here is that we multiply the volatility $\sigma$ by $\sqrt{X}$, which thus enforces nonnegative values of $X$ (or in the context of the CIR model, nonnegative interest rates $r_t$). This property is one that we will use later in Section \ref{sec:feller}.

\section{Measurement challenges}
\label{sec:measurement-challenges}

In order to apply the concepts outlined in Section \ref{sec:background}, we first need to obtain relevant data. In particular, fitting the Jones law of motion requires gathering data on both the output ($A$) and input ($I$) measures. While this might seem straightforward, there are thorny challenges in both accurately identifying $A$ and $I$, as well as obtaining data in practice. 
% The source of these difficulties is multifaceted, stemming from conceptual ambiguity about what constitutes meaningful proxies as well as empirical limitations in data availability and quality.

In this section, we elucidate the obstacles around properly specifying $A$ and $I$. For each measure, we examine the factors that engender uncertainty and impede accurate estimation. Our purpose is not to offer novel methodological solutions, as the problems appear largely intractable with current techniques. Rather, we aim to delineate the sources of uncertainty inherent in this endeavor, which can be incorporated into conclusions drawn from fitting the Jones law.

\subsection{The input measure might be unknown}
\label{sec:unknown-inputs}

To see what issues may arise, suppose that we are able to directly measure \( A \), and we know the Jones law of motion
\begin{equation} 
\frac{1}{A} \frac{dA}{dt} = \theta A^{-\beta} I^{\lambda} 
\end{equation}
holds (perhaps up to some noise) for \textit{some} input measure \( I \). The problem is that in many realistic situations, we do not actually know what the input measure \( I \) should be, and identifying it can be quite challenging. 

Here are a few specific ways in which misidentification of \( I \) could happen:
\begin{enumerate}

\item \textbf{Failure to account for alternative factors that influence $A$}. Suppose that $A$ is a measure of output for some domain of scientific R\&D, and $I$ is a proxy for the number of researchers in that domain. In this case, we might miss other factors such as spillovers from research in other adjacent domains, or the improvement of machines or relevant scientific equipment. For example, if \( A \) is a measure of software efficiency in a domain, we might miss that the scaling of computational resources used for experiments, in addition to $I$, also contributes to software progress.

\item \textbf{Input measures may be too narrow or broad}. Consider again that \( A \) is a parochial metric of efficiency, it is often not clear whether to use a ``narrow" or ``broad" measure for the inputs \( I \). For instance, if \( A \) represents the efficiency of computer chess engines, then we can make a case for both narrow input measures such as the number of people working on frontier chess engine projects, and for broad input measures such as the number of researchers working on game-playing programs worldwide. When a field is new, these differences are quantitatively significant: narrow input measures can often grow far faster than broad input measures because they start from a lower base.

\item \textbf{Not accounting for price effects when $I$ is large}. If \( I \) is a measure of spending, then it might be difficult to convert it to real inputs due to price impact effects. This becomes more relevant when substantial resources are already being spent to increase \( A \), e.g. for TFP of an economy or Moore's law in hardware efficiency. A doubling of spending does not necessarily mean we get twice the effective research input. It is possible to correct for this to some extent, e.g. by dividing spending measures by estimates of researcher wages as is done by \cite{bloom2020ideas}, but as \cite{ekerdt2023} argue, even this might fail to adequately control for e.g. the marginal researcher not having the same productivity as the average researcher.

\item \textbf{Changes in the meaning of ``patents" under different legal regimes}. Suppose \( I \) is estimated by looking at the number of patents filed in a domain, then it might be strongly affected by changes in legal regimes of intellectual property that do not necessarily have much to do with R\&D. If we were to use such data for the research inputs, the lack of sensitivity of \( A \) to \( I \) would lead us to estimate very low values for \( \lambda \). Moreover, patents are frequently also used as a measure of research output, and this same critique applies in that case. 

\end{enumerate}

\cite{bloom2020ideas} were not unaware of these difficulties. In fact, this obstacle appears to have troubled them significantly, as they explain in the quoted passage below:

\begin{quote}
Our selection of cases is driven primarily by the requirement that we are able to obtain data on both the “idea output” and the corresponding “research input.” We looked into a large number of possible cases to study, only a few of which have made it into this paper; indeed, we wanted to report as many cases as possible... However, it proved impossible to get a series for the research input that we felt corresponded to the idea output. For example, the Nordhaus price of light series would make a great additional case. But many different types of research contribute to the falling price of light, including the development of electric generators, the discovery of compact fluorescent bulbs, and the discovery of LEDs. We simply did not know how to construct a research series that would capture all the relevant R\&D. The same problem applies to the other cases we considered but could not complete... In the end, we report the cases in which we felt most confident. 
\end{quote}

The problem of finding a good input measure \( I \) remains challenging, with no easy solutions. Ideally, we would let our choice of \( I \) be informed by the data: the input measure we ought to use is the one that gives a stochastic version of the Jones law of motion its best fit with the data we have for \( A \). Unfortunately, in many practical situations, this method turns out to be far too optimistic.

For instance, rejecting the hypothesis \( \lambda = 0 \) for the Jones law of motion using some measure of inputs \( I \) should be a basic threshold for any serious input measure to exceed before it is relied upon for predictions. In practice, however, it is often the case that \textit{no} input measures significantly\footnote{In a statistical sense; e.g. in bootstrapping, using a likelihood ratio test, using model selection criteria, etc.} improve the law of motion's fit with data. Since we are frequently unable to even beat the \( \lambda = 0 \) baseline, the hope that we can pick the right measure of inputs by relying on empirical evidence over priors is often futile. We will go into greater detail about this in Section \ref{sec:case-studies}.

\subsection{The output measure might be unknown}
\label{sec:unknown-outputs}

This is the mirror image of the problem from \hyperref[sec:unknown-inputs]{the previous section}: we are able to directly measure \( I \), and we know the Jones law of motion
% \begin{equation} 
% \frac{1}{A} \frac{dA}{dt} = \theta A^{-\beta} I^{\lambda} 
% \end{equation}
holds for some \( A \), but we are not sure how to measure or construct a proxy for $A$. This problem comes up most frequently when \( A \) is a latent variable inferred from the data using a model, rather than being directly measured, as in this case model uncertainty can impact what we think the past trajectory of \( A \) has been. Most measures of ``efficiency" are latent variables, meaning that this problem is one that is encountered often; but it is more serious in some domains than in others.

Perhaps the most salient example in this category in growth economics is when \( A \) has something to do with a measure of TFP. As TFP is a latent variable of growth models, it is not directly inferred from the data and is sensitive to changes in model specification. Estimates of TFP, therefore, can depend on a wide range of model properties:

\begin{enumerate}
    \item \textbf{The factors that are included in the model.} For instance, \cite{jones2022past} includes a ``misallocation term" \( M \) in the growth model used to estimate TFP (which is denoted in the paper by \( Z \)), a term that is meant to capture variation in output at a fixed level of ``physical technology" due to resource misallocation. The inclusion of this factor leaves less growth to be explained by the stock of ideas \( A = Z/M \), which means we would estimate a slower growth in \( A \) and thus lower returns to research \( r \) for a fixed input time series \( I \).
    
    \item \textbf{How human capital is estimated.} Many sources do this by taking country-wide measures of educational attainment from datasets, such as \cite{barro2013new}, and combining them with estimates of returns to schooling: a particular source that follows this approach to produce worldwide TFP estimates is \cite{feenstra2015next}. In this case, TFP estimates can be influenced strongly by how much GDP growth we attribute to human capital versus how much is left over for TFP to explain.

    \item \textbf{How the factor shares in the economy are estimated.} A standard practice, followed by \cite{feenstra2015next} to a first approximation, is to match the labor elasticity of output with the factor share of labor in the economy, estimated by dividing the total amount of wage payments in the economy by GDP. The capital share is then estimated by assuming joint constant returns to scale to both labor and capital so that the two associated factor elasticities must sum to \( 1 \). If additional forms of labor compensation are missed, if other significant factors of production (land in agrarian economies, natural resources in some other economies, etc.) are neglected, or if some factors in the economy have substantial market power; estimates of TFP obtained in this way could be biased.

    \item \textbf{How real GDP is estimated.} This seems like a strange point to include in this list, and for countries such as the US there is comparatively little controversy about economic data, but real GDP data is significantly disputed for countries like China. \cite{bosworth2008accounting} uses official data on Chinese real GDP growth and estimates that Chinese TFP grew by \( 3.6 \%/\text{year} \) from 1978 to 2004 with human capital taken into account, while \cite{feenstra2015next} uses a real GDP series which attributes more real output to China in 1978, resulting in TFP growth estimates in the vicinity of \( 0 \%/\text{year} \). It is beyond the scope of this paper to weigh in on this debate, but we think it is worth noting just how substantial our uncertainty can be when it comes to this point.
\end{enumerate}

\cite{erdil2023china} elaborates further on how the TFP estimates from \cite{feenstra2015next} might change if these factors are taken into account. The changes are substantial for many countries and can make the difference between TFP remaining flat as opposed to showing sustained growth over long periods of time. This is problematic, as the exact value of the returns to research \( r \) is of interest in many cases, and the rough approximation from dividing the growth rates in outputs and inputs shows how overestimating or underestimating the output growth rate can lead to inaccurate estimates of \( r \).

Besides TFP, similar problems occur for other latent variables. For example, in the case of software R\&D, we often want to find a multiplicative metric of ``software efficiency" that changes in a particular domain over time, just as TFP is a measure of ``resource use efficiency" with the same character. As this variable is not directly measured, it is also a latent variable. However, turning it into a single-dimensional efficiency multiplier is difficult because of the multidimensional and scale-dependent nature of software progress. Software innovations could lead to greater compute savings for larger applications than smaller ones, e.g. by improving the complexity class of a particular problem; and they could be heterogeneous across different problems in the same domain, e.g. a narrow benchmark might become easier to beat with fewer resources while a broader benchmark sees less progress.

We consider machine learning as a concrete example of a domain where previous work has attempted to measure the extent of software R\&D progress. Here, existing work generally follow one of two approaches. They either fix a specific benchmark and performance threshold and measure how the resources needed to attain that level of performance decrease over time (\cite{hernandez2020measuring}), or they develop a predictive model of model performance based on resource inputs that makes simplifying assumptions to be able to produce a single ``averaged" quantity \( A \) that measures software efficiency (\cite{erdil2022algorithmic}). Neither approach is completely satisfactory, but both should be preferable to having no information about software efficiency at all.

\begin{comment}
\subsection{The idea production function might not be Cobb-Douglas}

Another possibility is that \( A \) satisfies a more general law of motion given by

\begin{equation} \frac{dA}{dt} = f(A, I) \end{equation}

for some function \( f: (\mathbb R^{+})^2 \to \mathbb R^{+} \), but this function is not Cobb-Douglas, i.e. not of the form \( \propto A^{\phi} I^{\lambda} \). In this case, values of \( \beta, \lambda \) that we estimate will still give us information about the local input elasticities of the idea production function \( f \), but they won't necessarily imply that e.g. an \( AK \) model will exhibit a singularity whenever \( r + \alpha > 1 \), as \( r \) would no longer be a constant.
\end{comment}

%\setcounter{equation}{0}

\section{Estimation strategies}
\label{sec:estimation-strategies}

As we alluded to in Section \ref{sec:prior-work}, the challenges to estimating the idea production function extend beyond just measurement challenges. In particular, another crucial consideration is the methods we use to obtain parameter estimates given some data. In this section, we will thus assume that the \hyperref[sec:measurement-challenges]{measurement challenges} have already been addressed, and discuss how we might go about estimating the model parameters given some amount of data.

As the Jones law of motion has three parameters—$\lambda$, $\beta$ and $\theta$—we need at least three time periods over which we know the growth rate of \( A \) and the input series \( I \) for identification. When the law of motion also has a nontrivial noise structure, the requirements go up further: a simple homoskedastic noise term raises the number of data points needed to \( n = 4 \), and more general noise structures require even more parameters for identification.

If we have many more than four data points, then the estimation of the Jones law of motion can proceed according to standard frequentist methods such as ordinary least squares (OLS) regression or maximum likelihood estimation (MLE). If not, we have to use prior information to constrain the parameter values to some extent to get anything out of the equation.

Table \ref{tab:summary-table-strategies} summarizes the methods we discuss in this section. If the reader is uninterested in the technical details of these methods, we recommend reading Table \ref{tab:summary-table-strategies} and skipping the rest of this section to go to Section \ref{sec:case-studies}, where we report the results of applying the methods we discuss here in concrete situations.

\begin{table}[h]
\centering
\small
\begin{tabularx}{\textwidth}{>{\hsize=0.5\hsize}X X X >{\hsize=1.2\hsize}X}
\toprule
Method & Summary & Conditions for Use & Accuracy \\
\midrule
Naive method & Estimate \( r = g_A/g_I \). & The average growth rates of the inputs and outputs must be known. & Appropriate for back-of-the-envelope calculations, order of magnitude estimates, etc. \\
\addlinespace
Linear regression & Approximate the law of motion log-linearly and estimate \( \lambda, \beta \) by linear regression methods. & \( A \) must be strictly increasing and we should have sufficiently high-frequency data for the linear approximation to be valid. & Decent performance when the conditions are met. If model uncertainty is not substantial, MLE should be preferred for efficiency reasons. \\
\addlinespace
MLE & Write down a stochastic generalization of the law of motion and estimate its parameters using MLE. & If the likelihood function is approximately evaluated, then high-frequency data may be needed for the approximation to be valid. We must also have enough data for the MLE problem to admit a unique solution. & Best performance when we have a good model and the quantity of data to support it. If model uncertainty is significant, linear regression might be preferred. \\
\addlinespace
Bayesian methods & Write down a stochastic generalization of the law of motion and perform a Bayesian update on a prior over the parameters using the implied likelihood function. & If the likelihood function is approximately evaluated, then high-frequency data may be needed for the approximation to be valid. & Reduces to MLE if data quantity is sufficiently large. If data is scarce, it is better than the naive method because it allows for quantification of uncertainty if the stochastic model is not too badly specified. \\
\bottomrule
\end{tabularx}
\caption{Summary of the different methods discussed in this section.}
\label{tab:summary-table-strategies}
\end{table}

\subsection{Solving the differential equation}
\label{sec:solving-differential-eqn}

To set up our estimation approaches, we will require an explicit, closed-form solution of the differential equation defined by the Jones law of motion. We thus begin by computing this solution. This is a simple calculation: starting from 
\begin{equation} 
\frac{1}{A} \frac{dA}{dt} = \theta A^{-\beta} I^{\lambda}, 
\end{equation}
we separate variables to obtain
\begin{equation} \
\frac{1}{\beta} \frac{d A^{\beta}}{dt} = A^{\beta - 1} \frac{dA}{dt} = \theta I^{\lambda} 
\end{equation}
and integrate both sides from \( t = t_1 \) to \( t = t_2 \), which yields:
\begin{equation}
\label{eq:solution}
    \frac{A(t_2)^{\beta} - A(t_1)^{\beta}}{\beta} = \theta \int_{t_1}^{t_2} I(t)^{\lambda} \, dt = \theta I_\lambda(t_1, t_2)^{\lambda},
\end{equation}
where we have made the following definition for notational convenience:
\begin{equation}
I_{p}(t_1, t_2) = \left( \int_{t_1}^{t_2} I(t)^{p} \, dt. \right)^{1/p} 
\end{equation}
One immediate conclusion is that each pair \( (A(t_1), A(t_2)) \) defines one equation between the parameters of the model, assuming that the norms \( I_{\lambda}(t_1, t_2) \) are known for all values of \( \lambda \), which under reasonable assumptions is equivalent to knowing the distribution of \( I \) as a random variable over the time interval \( [t_1, t_2] \). We therefore need at least three such pairs to identify the parameters of the law of motion, which means we must have at least four data points at which we know the value of \( A \).

\subsection{The naive method: dividing the growth rates}
\label{sec:naive-method}

% As mentioned in the introduction, the simplest way to estimate the parameter \( r \) is by the ratio \( g_A/g_I \). This estimate works in an exponential growth equilibrium and requires very little data about \( A \) or \( I \) to be calculated,

We now turn our attention to the first parameter estimation strategy that we consider. As mentioned in Section \ref{sec:jones-law}, the simplest way to estimate the returns to R\&D $r$ is by the ratio of growth rates of outputs to inputs. This approach works best in an exponential growth equilibrium and requires very little data about \( A \) or \( I \) to be calculated. Indeed, in such an equilibrium the left-hand side of the equation is constant, so the right-hand side must be constant as well. Taking logarithms and differentiating with respect to time immediately gives
\begin{equation}
    \lambda \times \frac{\dot I}{I} = \beta \times \frac{\dot A}{A},
\end{equation}
or equivalently,
\begin{equation}
    r = \frac{\lambda}{\beta} = \frac{\dot A/A}{\dot I/I} = \frac{g_A}{g_I}.
\end{equation}
Here, \( g_A, g_I \) denote the growth rates of \( A \) and \( I \) respectively. This relation is often used for naive estimates of \( r \) in contexts where little data is present or a rough back-of-the-envelope calculation is thought to be sufficient.\footnote{This is done in \cite{davidson2023compute}, and roughly but not exactly what is done in \cite{bloom2020ideas} (see \ref{sec:diminishing-returns}).} 

We might also be interested to know what happens if we are not exactly in such an equilibrium, or if we follow a noisy version of the Jones law of motion instead. To this end, we can obtain some useful results from \hyperref[sec:solving-differential-eqn]{the explicit solution} to the differential equation. Fixing a reference time \( t_0 \) and two future times \( t_1, t_2 \), we know that the equality
\begin{equation}
\label{eq:ratio-solution}
    \left( \frac{(A(t_2)/A(t_0))^{\beta} - 1}{(A(t_1)/A(t_0))^{\beta} - 1} \right)^{1/\lambda} = \frac{I_{\lambda}(t_0, t_2)}{I_{\lambda}(t_0, t_1)}
\end{equation}
must hold: this follows immediately upon evaluating the explicit solution from Equation \ref{eq:solution} at times \( t_2, t_0 \) and \( t_1, t_0 \), then dividing the resulting two equations. It can be shown that if $I$ is continuous and increasing, and we consider the limit of large $\lambda$,\footnotemark we have 
\begin{equation}
\label{eq:asymptotic-naive-method}
    \lim_{\lambda \to \infty} r(\lambda) = \lim_{\lambda \to \infty} \frac{\lambda}{\beta(\lambda)} = \frac{\log \left( \frac{A(t_2)}{A(t_1)} \right)}{\log \left( \frac{I(t_2)}{I(t_1)} \right)}.
\end{equation}
\footnotetext{We leave the details of this derivation to Appendix \ref{sec:derivation-naive-method}.}
As the right-hand side is exactly the ratio of the mean growth rate of \( A \) to that of \( I \), Equation \ref{eq:asymptotic-naive-method} says that the naive method of dividing the growth rates works well when the inputs \( I \) are monotonic and the true value of \( \lambda \) is suitably large. These conditions are usually much easier to meet than \( I \) and \( A \) both growing exponentially, but the naive method can still mislead us about the value of \( r \) when \( \lambda \) is small, as we will see later in the \hyperref[sec:case-studies]{case studies section}.

The above calculations also suggest refinements of the naive method to cases where we have information about the value of \( \lambda \). For instance, if we know the exact value of \( \lambda \), then we can directly use the relation
\begin{equation}
\label{eq:ratio-identity}
    \frac{(A(t_2)/A(t_0))^{\beta} - 1}{(A(t_1)/A(t_0))^{\beta} - 1} = \left( \frac{I_{\lambda}(t_0, t_2)}{I_{\lambda}(t_0, t_1)} \right)^{\lambda}.
\end{equation}
to estimate \( \beta \). Solving this for the value of \( \beta \) requires us to know both ratios \( A(t_2)/A(t_0) \) and \( A(t_1)/A(t_0) \), but if \( A(t_1)^{\beta} \gg A(t_0)^{\beta} \), the LHS will be well-approximated by \( (A(t_2)/A(t_1))^{\beta} \), so we will obtain
\begin{equation} \left( \frac{A(t_2)}{A(t_1)}\right)^{\beta} \approx \left( \frac{I_{\lambda}(t_0, t_2)}{I_{\lambda}(t_0, t_1)} \right)^{\lambda} 
\end{equation}
which gives us a modified form of the naive method upon taking logarithms where the instantaneous inputs \( I \) are replaced by appropriate \( L^{\lambda} \) norms. If \( \lim_{t \to -\infty} A(t) = 0 \), this approximation becomes an exact equality upon passing to the limit \( t_0 \to -\infty \), but in practice this limit is of little significance. This is because domains where \( \lim_{t \to -\infty} A(t) = 0 \) holds and where we expect the Jones law of motion with the current parameters to have held indefinitely further back into the past are scarce. In practice, we are better off using a reference point \( t_0 \) such that  \( A(t_1)^{\beta} \gg A(t_0)^{\beta} \), though it can once again be difficult to know which values of \( t_0 \) are small enough that we should expect this, as we do not know the value of \( \beta \).\footnote{A notable variant of this method is used in \cite{bloom2020ideas}. Since we do not recommend using this approach, we direct the reader to Appendix \ref{sec:diminishing-returns} for more details.}

\subsection{Stochastic laws of motion}
\label{sec:randomness}

If the naive approach is only good for obtaining rough estimates of $r$, what alternatives can we use? In this section we present a range of approaches based on generalizing the Jones law of motion to a stochastic model.
% that defines a probability distribution over the output time series \( A \) conditional on the inputs \( I \). 
In particular, recall the core solution to the Jones law of motion:
\begin{equation}
    A(t_2)^\beta - A(t_1)^\beta = \beta \theta \int_{t_1}^{t_2} I(t)^\lambda \: dt.
\end{equation}
The core idea is to determine a stochastic relationship between the outputs $A$ and $I$ that's analogous to this deterministic relationship. In particular, the input intensity \( I(t) \) effectively acts as a time scaling factor in the integral on the right hand side, such that in time \( dt \) we make an amount of progress proportional to \( I(t)^{\lambda} \) if progress is measured by how much \( A(t)^{\beta} \) increases over the time interval. We want the stochastic generalization to satisfy the same property in expectation\footnote{It's rather unclear what this means, as taking expectations doesn't commute with raising \( A \) to the power \( \beta \) on the left hand side: in general, \( \mathbb E[A^{\beta}] \neq \mathbb E[A]^{\beta} \). The choice of which expectation to match to the deterministic law affects how we try to generalize to the stochastic case.}, and have a "natural" noise structure in an as of yet unclear sense.

In this section we consider four different ways of generalizing the Jones law of motion to a stochastic setting. These stochastic methods have the advantage that they naturally lend themselves to traditional frequentist (e.g. maximum likelihood) or Bayesian methods of estimation, since they come together with likelihood functions on observations depending on parameters that can be computed (at least in principle).

Some complications that are absent in the deterministic case can arise when we attempt to do this, as separating variables in Stochastic Differential Equations (SDEs) is not as straightforward unless the variance structure of the noise term is well-behaved. In any specific situation, the choice of noise structure should ultimately be based on the data; but considerations in this section can inform such a choice, as well as help avoid overfitting the noise structure to the data.

\subsubsection{Flexible Lévy estimation procedure}
\label{sec:first-proposal}

The first idea is to treat each unit of R\&D input independently, sampling different research productivities. Let \( X_p(\omega) \) be a Lévy process with parameters \( p \) normalized such that \( X_p(0) = 0 \), and define

\begin{equation}
    \label{eq:stochastic-first-proposal}
    A(t_2)^{\beta} - A(t_1)^{\beta} = \beta \cdot X_p \left( \int_{0}^{t_2} I(t)^{\lambda} \, dt \right) - \beta \cdot X_p \left( \int_{0}^{t_1} I(t)^{\lambda} \, dt \right)
\end{equation}

for times \( t_1 < t_2 \). When the Lévy process \( X_p \) is a deterministic pure drift process \( X_p(\omega) = \theta \cdot \omega \), this simplifies to the deterministic Jones law of motion. The advantage of this specification is that it is a closed-form for the process \( A \), so we avoid having to approximate the solutions to a stochastic differential equation that has no closed-form solution.

Here we choose a general Lévy process as opposed to a drift-diffusion process with no jump component to model software efficiency time series which exhibit discontinuities. Choosing \( X_p \) to be a drift-diffusion process can lead to biased estimates of the coefficients in these cases. A Lévy process can also better model skewness in the increments, while a drift-diffusion process would force the distribution's mean and median to coincide. If working with a time series that does not have such behavior, one can always restrict \( X_p \) to be of drift-diffusion form or pick a model class for \( X_p \) which includes all such processes within, so that an optimizer can find them if they indeed have the best fit with data.

A serious problem with this estimation procedure arises when \( X_p \) is chosen to be a Lévy process with a strictly positive probability of decreasing over some input interval. Then regardless of which value of \( A \) we start the process from, there will be a strictly positive chance that the process defining \( A^{\beta} \) will fall below zero, giving us no well-defined value for \( A = (A^{\beta})^{1/\beta} \). As a consequence, strictly speaking, this does not define a law for \( A \) unless \( X_p \) is an almost surely non-decreasing Lévy process such as a gamma or Poisson process. We can choose \( X_p \) to be non-decreasing, but then we lose our ability to fit time series of \( A \) which exhibit local decreasing behavior, e.g. TFP time series. 

When \( X_p \) is drift-diffusion, there is a way to use Feller diffusion to avoid this problem, which we outline in Section \ref{sec:feller}. It involves scaling down the Lévy process the closer \( A \) gets to zero, thus avoiding the possibility of \( A \) becoming negative. This also readily generalizes to Lévy processes whose jump component is almost surely increasing, e.g. a stable process with maximal skewness parameter, as explained in \cite{nolan2020univariate}. However, this involves modifying the law of motion in a way that makes it less tractable to solve. If we intend to stick to Equation 
\ref{sec:first-proposal}, we can ensure positivity by defining a latent process \( H \) following 

\begin{equation}
    \label{eq:stochastic-first-proposal-latent}
    H(t_2) - H(t_1) = \beta \cdot X_p \left( \int_{0}^{t_2} I(t)^{\lambda} \, dt \right) - \beta \cdot X_p \left( \int_{0}^{t_1} I(t)^{\lambda} \, dt \right)
\end{equation}

and then set \( A(t) = \max(H(t), 0)^{1/\beta} \). This ensures that \( A \) is always well-defined. In practice, we do not need to pay much attention to this technical condition, as empirical estimates often find processes \( X_p \) that have a vanishingly small chance of ever hitting zero. However, we include it to ensure that what we are doing makes sense. We will omit this technical condition from later subsections, but they should formally be taken to be about a latent process \( H \) instead of about \( A \) directly because of this positivity constraint.

\subsubsection{Synchronized input stochastic estimation}
\label{sec:second-proposal}

A similar but slightly different structure is to consider inputs being invested at the same time, sampling the \emph{same} research productivity $dX_p$. This is given by
\begin{equation}
\label{eq:stochastic-second-proposal}
    A(t_2)^{\beta} - A(t_1)^{\beta} = \beta \cdot \int_{t_1}^{t_2} I(t)^{\lambda} \, dX_p,
\end{equation}
where once again $\theta$ has been absorbed into $X_p$. We can compare this to the previous model in Section \ref{sec:first-proposal}. The lack of same-time correlation in research productivity for the previous model means that the causal influence of the inputs on \( A(t_2) \) conditional on \( A(t_1) \) factors through \( I_{\lambda}(t_1, t_2) \), just like with the deterministic law; but this does not happen for the model in this section.

A useful case to see the difference between these two laws of motion for \( A \) is to focus on the case when \( X_p \) is a stable process, i.e. its increments are stable distributed (see \cite{borak2005stable} for a reference on such distributions) with stability, skewness, rate, and scale parameters \( \alpha_p ,\beta_p, \mu_p, c_p \) respectively. In this case, it is straightforward to see the following:
\begin{align}
     X_p \left( \int_{0}^{t_2} I(t)^{\lambda} \, dt \right) - X_p \left( \int_{0}^{t_1} I(t)^{\lambda} \, dt \right) &\sim \text{Stable}\left(\alpha_p, \beta_p, \mu_p \cdot I_{\lambda}(t_1, t_2)^{\lambda}, c_p \cdot I_{\lambda}(t_1,t_2)^{\lambda/\alpha} \right) \\
     \int_{t_1}^{t_2} I(t)^{\lambda} \, dX_p &\sim \text{Stable}\left(\alpha_p, \beta_p, \mu_p \cdot I_{\lambda}(t_1, t_2)^{\lambda}, c_p \cdot I_{\alpha \lambda}(t_1,t_2)^{\lambda/\alpha} \right).
\end{align}
The only difference turns out to be whether the norm \( I_{\lambda} \) or \( I_{\alpha \lambda} \) appears in the scale term for the distribution. While this may seem like a small difference, it is theoretically significant and affects estimates of \( \lambda, \beta, r \) substantially in practice. The exact noise structure chosen is therefore of considerable importance.

\subsubsection{Scale-invariant stochastic estimation}
\label{sec:third-proposal}

A third proposal arises from wanting \( dA/A \) to satisfy a condition of scale invariance, akin to the interest rate in the CIR model (see Section \ref{sec:special-cases}). Let us first bring back the Jones law of motion 
\begin{equation} 
\frac{d A^{\beta}}{A^{\beta}} = \theta \beta A^{-\beta} I^{\lambda} \, dt \end{equation}
Following the same scale invariance principle as in the CIR model, we should treat \( A^{-\beta} I^{\lambda} \) as a time scaling factor. This suggests, for \( X_p \) a Lévy process, the following stochastic generalization:
\begin{equation}
\label{eq:stochastic-third-proposal}
    \frac{d A^{\beta}}{A^{\beta}} = \beta \int_{0}^{A^{-\beta} I^{\lambda} \, dt} dX_p,
\end{equation}
where we absorb the constant $\theta$ into the Lévy process. 
This can be expressed conveniently when \( X_p \) is a stable process with stability, skewness, rate, and scale parameters \( \alpha_p ,\beta_p, \mu_p, c_p \) respectively as
\begin{equation} 
\frac{d A^{\beta}}{A^{\beta}} \sim \beta \cdot \text{Stable}\left(\alpha_p, \, \beta_p, \, \mu_p A^{-\beta} I^{\lambda} \, dt, \, c_p A^{-\beta/\alpha_p} I^{\lambda/\alpha_p} \, (dt)^{1/\alpha_p} \right).
\end{equation}
The dependence of the right-hand side on \( A \) makes this equation intractable to solve in closed form. However, we can approximate it at short time horizons by assuming that \( A \) is locally constant on the right-hand side, which yields
\begin{equation} 
\frac{A(t_2)^{\beta} - A(t_1)^{\beta}}{\beta} \stackrel{\text{approx}}{\sim} \text{Stable}\left(\alpha_p, \, \beta_p, \, \mu_p I_{\lambda}(t_1, t_2)^{\lambda}, \, c_p A^{\beta-\beta/\alpha_p} I_{\lambda}(t_1, t_2)^{\lambda/\alpha_p} \right) 
\end{equation}
as a good approximation when \( t_2 - t_1 \) is sufficiently small. As before, the same caveats about positivity mentioned in Section \ref{sec:first-proposal} should apply here, but we can lose these caveats in the approximation when the skewness parameter \( \beta_p \) is close to \( 1 \), which empirically turns out to be the case.

A more general version of this noise structure can change the exponent \( \beta - \beta/\alpha_p \) into a free parameter, but if we wish to avoid adding an additional free parameter to the model, setting this exponent equal to \( \beta - \beta/\alpha_p \) is theoretically justified by the above scale invariance argument. A choice of zero for this exponent can also be justified, following Equation \ref{eq:stochastic-first-proposal} instead of Equation \ref{eq:stochastic-third-proposal}. In the end, the decision of which exact noise structure to use should be an empirical matter, as theory does not set strong constraints.

\subsubsection{Feller diffusion}
\label{sec:feller}

As we alluded to in Section \ref{sec:first-proposal}, one practical issue that we might want to deal with is when $X_p$ has a strictly positive probability of decreasing over some input interval. One way to get around this is to consider a Feller diffusion process. 

In particular, in the special case of \( \alpha_p = 2 \), i.e. when we choose \( X_p \) to be a drift-diffusion process, we can use known explicit solutions to Feller diffusion to obtain a closed form for \( A(t_2) \) given \( A(t_1) \) if we use the noise structure proposed in Section \ref{sec:third-proposal}. We interpret this noise structure as adding a dependence on the inputs \( I \) to a Feller diffusion by subordination: we define \( A(t_2) = S(I_{\lambda}(t_1, t_2)^{\lambda}, A(t_1)) \) where \( S(t, S_0) \) is a solution of the stochastic differential equation
\begin{equation}
    \label{eq:third-proposal-gaussian-law-of-motion}
    \frac{dS}{S} = \theta S^{-\beta} \, dt + \sigma S^{-\beta/2} \, dW_t
\end{equation}
with boundary condition \( S(0) = S_0 \), which means \( A \) solves the SDE
\begin{equation}
    \label{eq:third-proposal-gaussian-law-of-motion-two}
    \frac{dA}{A} = \theta A^{-\beta} I^{\lambda} \, dt + \sigma A^{-\beta/2} I^{\lambda/2} \, dW_t.
\end{equation}
\cite{roodman2020probability} notes that the substitution \( X = S^{\beta} \) transforms Equation \ref{eq:third-proposal-gaussian-law-of-motion} into
\begin{align}
    dX &= dS^{\beta} = \beta S^{\beta-1} \, dS + \frac{\beta(\beta-1)}{2} S^{\beta-2} \, dS^2 \\
    dX &= \left(\beta \theta + \frac{\beta(\beta-1)}{2} \sigma^2 \right)\, dt + \beta \sigma S^{\beta/2} \, dW_t \\
    dX &= \left(\beta \theta + \frac{\beta(\beta-1)}{2} \sigma^2 \right)\, dt + \beta \sigma \sqrt{X} \, dW_t
\end{align}
upon appropriate use of Ito's lemma. Note that the bias term $\beta (\beta-1)/2$ coming from the second-order contribution to \( dt \) from Ito's lemma makes the role of \( \theta \) in this equation slightly different from the one found in Equation \ref{eq:stochastic-third-proposal}, but this is a relatively minor difference and amounts to a reparametrization of the drift term that does not affect the role of the important parameters \( \beta, \lambda \) in the model.

The final equation we obtain is a Feller diffusion in the variable \( X \), and the associated Fokker-Planck equation (which governs how the probability density of $X$ evolves over time) admits a closed-form probability density solution described in equation 42 of \cite{roodman2020probability}. Combining these, we can recover a closed-form solution for the forward probability density of Equation \ref{eq:third-proposal-gaussian-law-of-motion}. This is not very useful when we are working with high-frequency data, as in that regime naive approximations tend to be suitably good, but it will be very helpful when we want to perform Bayesian updates using only low-frequency data.

There is a remaining important technical condition that should be noted. For our \( X \) to correspond to a solution \( S \) of the original untransformed equation, we must have that \( X \) is almost surely positive: as otherwise we cannot raise it to a fractional power \( 1/\beta \). This condition requires the drift coefficient \( \beta \theta + \frac{\beta(\beta-1)}{2} \sigma^2 \) to be strictly positive, which is equivalent to asking for \( \theta > \sigma^2 (1-\beta)/2 \). This restriction is vacuous when \( \beta \geq 1 \), but in the interval \( 0 < \beta < 1 \) it places constraints upon admissible parameter values. This should be taken into account when using the law of motion from Equation \ref{eq:third-proposal-gaussian-law-of-motion-two}.

\subsection{Bayesian inference methods}
\label{sec:bayesian}

Once we have a stochastic law of motion, such as the ones presented in Section \ref{sec:randomness}, we can attempt to deal with domains where data is scarce using Bayesian methods. Specifically, given a noise structure with parameters \( \vec{p} \), we can choose a prior over them and then perform a Bayesian update on this prior based on the data we observe. This works even if the model is underidentified, i.e. if we have fewer data points than we have parameters.

Suppose that we have a Markov process for \( A \) with forward probability densities \( f_I(A_{t_2}, A_{t_1}, t_2, t_1, \vec{p}) \) which give the density that \( A(t_2) = A_{t_2} \) at time \( t_2 \) given the inputs \( I \), the model parameter vector \( \vec{p} \), and the value \( A_{t_1} \) of \( A \) at time \( t_1 \). Then, the likelihood we assign to the collection of pairs \( D = \{ (t_1, A_{t_1}), (t_2, A_{t_2}), \ldots, (t_n, A_{t_n}) \} \) is given by the product
\begin{equation} L(\vec{p}) = \prod_{k=1}^{n-1} f_I(A_{t_{k+1}}, A_{t_k}, t_{k+1}, t_k, \vec{p}) \end{equation}
and the associated log-likelihood is therefore given by
\begin{equation} \mathcal L(\vec{p}) = \sum_{k=1}^{n-1} \log f_I(A_{t_{k+1}}, A_{t_k}, t_{k+1}, t_k, \vec{p}). \end{equation}
In maximum likelihood estimation, we pick the value of the parameters in the parameter vector \( \vec{p} \) to maximize \( \mathcal L(\vec{p}) \). For Bayesian inference, we instead take a prior \( \mathbb P(\vec{p}) \) over the parameters \( \vec{p} \) and compute the posterior by the Bayes rule:
\begin{equation} 
\mathbb P(\vec{p} \vert D) = \frac{\mathbb P(D \vert \vec{p}) \mathbb P(\vec{p})}{\mathbb P(D)} = \frac{L(\vec{p}) \mathbb P(\vec{p})}{\mathbb P(D)}.
\end{equation}
In practice, due to the intractability of computing \( \mathbb P(D) \) it is convenient to use Markov Chain Monte Carlo (MCMC) methods that can sample from the posterior without the need to compute the normalization factor \( \mathbb P(D) \). In this paper, we use Python's PyMC library from \cite{abrilpla2023pymc} for MCMC inference. Experimentally, we noticed that Hamiltonian-based samplers such as the NUTS sampler from \cite{hoffman2014no} could struggle when there is not much data to update on, and more inefficient methods that do not run a risk of divergence such as the differential evolution (DE) Metropolis sampler can be more effective.

The choice of the prior \( \mathbb P(\vec{p}) \) can also be important, and in general we recommend fairly uninformative choices such as Cauchy or half Cauchy priors over dimensionless parameters in domains we do not have much information. It is important to not sneak in our intuitive beliefs that might originate from knowing the data into the prior, as this would lead to an inefficient ``double update" on the available evidence \( D \). Deviation from this recommendation should only be considered in specific cases where we have good independent reasons for our priors to be narrower. In that case, it might be more principled to start with an uninformative prior and also incorporate those reasons into our Bayesian inference.

\subsection{Approximate linear regression}
\label{sec:linear-regression}

Though the discussion from Section \ref{sec:randomness} is important for fitting good models in practice, they are ultimately rather sophisticated, especially when general Lévy processes are used. Since MLE does not guarantee consistency unless the noise structure is chosen correctly, it is always tempting to find some way of estimating the law of motion by using linear regression. While there is no exact way to do this estimation, in some situations we can approximate the Jones law of motion in a suitable way for linear regression to be applicable. Even if we do not use these approximations in practice, thinking of the problem of estimation in these terms allows us to rephrase some obstacles to getting good estimates in more standard language.

The essential ingredient in this approximation is Equation \ref{eq:solution-jones-form}, derived in the appendix, and repeated below for convenience.
\begin{equation}
\tag{\ref{eq:solution-jones-form} revisited}
    Q_{\beta}(t_1, t_2) = \frac{\left( \frac{A(t_2)}{A(t_1)} \right)^{\beta} - 1}{\beta} = \theta A(t_1)^{-\beta} I_\lambda(t_1, t_2)^{\lambda}.
\end{equation}
As \( \log Q_{\beta}(t_1, t_2) \approx \log \log (A(t_2)/A(t_1)) \) to top order\footnote{In fact this understates how good of an approximation this tends to be in practice, because the top term of the error also tends to have less variance than we might expect, so most of the bias affects estimation of \( \theta \) more than \( \beta, \lambda \). See Section \ref{sec:diminishing-returns} for more on this.}, we can naively get an approximation
\begin{equation}
\label{eq:linear-regression-approx}
    \log \log \left( \frac{A(t_2)}{A(t_1)} \right) \approx \log \theta - \beta \log A(t_1) + \lambda \log I_{\lambda}(t_1, t_2).
\end{equation}
This is almost in the right form for linear regression, but not exactly, as \( I_{\lambda}(t_1, t_2) \) is a function of \( \lambda \). If the value of \( \lambda \) is assumed to be known, this is not a problem. If it is not known, then we want to make the sampling period \( t_2 - t_1 \) sufficiently small so that the variance of \( I \) is low, and use a suitable approximation to \( I_{\lambda} \). A useful second-order expansion in this context is
\begin{align}
\log(I_{\lambda}(t_1, t_2)^{\lambda}) &= \log \left( \int_{t_1}^{t_2} I(t)^{\lambda} \, dt \right) \\
&= \log(t_2 - t_1) + \log \mathbb E_{t \sim (t_1, t_2)}[I^{\lambda}] \\
&\approx \log(t_2 - t_1) + \lambda \log \mathbb E_{t \sim (t_1, t_2)}[I] + \log \left( 1 + \frac{\lambda(\lambda-1)}{2} \frac{\operatorname{var}_{t \sim (t_1, t_2)}(I)}{\mathbb E_{t \sim (t_1, t_2)}[I]^2} \right).
\end{align}
The intuition is that we can drop the third term whenever \( \operatorname{var}_{t \sim (t_1, t_2)}(I) \ll \mathbb E_{t \sim (t_1, t_2)}[I]^2 \), which happens for \( t_2 - t_1 \) small e.g. whenever \( I \) is continuous and strictly positive. We then recover a regression of the form
\begin{equation}
\label{eq:linear-regression-approx-2}
    \log \left( \frac{1}{t_2 - t_1} \log \left( \frac{A(t_2)}{A(t_1)} \right) \right) \approx \log \theta  - \beta \log A(t_1) + \lambda \log \left( \frac{I_{1}(t_1, t_2)}{t_2 - t_1} \right) + \varepsilon_{t_1, t_2}.
\end{equation}
This is a model we can actually fit using standard linear regression methods such as ordinary least squares (OLS), as all the unknown parameters appear as intercepts or linear coefficients. When it works, it is often a useful first-pass method to use on any data, as more sophisticated methods can be harder to implement correctly and this can serve as a useful benchmark to compare the results of better models against.

\subsubsection{The output series must be strictly increasing}
\label{sec:regression-increasing}

An important caveat is that this noise structure \textit{forces} \( A \) to be increasing. This is, of course, true of the deterministic law; but need not hold for more general noise structures such as those from Section \ref{sec:randomness}, and it might also not hold in practice if \( A \) is taken to be a TFP time series, for instance. Even the case where \( A \) merely fails to be strictly increasing with positive inputs is problematic, because if \( A \) is constant over some time interval then the left-hand side will be \( - \infty \) while the right-hand side will be some finite value, ignoring the noise term.

Unfortunately, it is not clear how to repair the method so it generalizes to this case. Picking \( t_1, t_2 \) such that \( t_2 - t_1 \) is always big enough for \( A(t_2) > A(t_1) \) is sometimes good enough to get some results out of the method, but it is rather unprincipled and in tension with the need to make \( t_2 - t_1 \) small to ensure \( \operatorname{var}_{t \sim (t_1, t_2)}(I) \ll \mathbb E_{t \sim (t_1, t_2)}[I]^2 \).

\subsubsection{Multicollinearity can make estimation difficult}
\label{sec:multicollinearity}

Another problem that becomes apparent when the estimation process is cast in linear regression form is \textit{multicollinearity}: insofar as \( \log A \) and \( \log I \) are linearly correlated, this correlation will result in the covariance matrix \( \Sigma_{\log A(t_1), \log I_1(t_1, t_2)} \) having at least one small positive eigenvalue, which will become a large eigenvalue when the covariance matrix is inverted to find the OLS standard errors for \( \beta, \lambda \). The worst situation is if \( \log A \) and \( \log I \) are perfectly correlated: this happens when both of them grow exponentially, and in this case, all the information the regression can give us is that we should estimate \( r = \lambda/\beta \approx g_A/g_I \). We get no specific information about \( \beta \) or \( \lambda \) as individual parameters beyond that. Note that this problem is not exclusive to linear regression: the setting of OLS estimation simply provides a convenient illustration.

If \( \log A \) and \( \log I \) are correlated with some correlation coefficient \( 0 < \rho < 1 \), then a good rule of thumb is that our effective sample size for identifying both parameters separately (rather than identifying \( r = \lambda/\beta \), for instance) goes down relative to the \( \rho = 0 \) case by a factor equal to \( 1/(1 - \rho^2) \). This can be seen by examining the diagonal entries of the inverse correlation matrix
\begin{equation} \rho_{\log A, \log I}^{-1} = \begin{bmatrix} 1 & \rho \\ \rho & 1 \end{bmatrix}^{-1} = \frac{1}{1 - \rho^2} \begin{bmatrix} 1 & -\rho \\ -\rho & 1. \end{bmatrix} \end{equation}
as the OLS standard error covariance matrix of the coefficients will be given by \( n^{-1} \rho_{\log A, \log I}^{-1} \sigma_{\varepsilon}^2 \) where \( n \) is the sample size and \( \sigma_{\varepsilon}^2 \) is the residual variance. When we change the correlation from \( 0 \) to a positive value \( \rho \), the diagonal entries of \( \rho_{\log A, \log I}^{-1} \) are divided by \( 1 - \rho^2 \), and hence \( n \) must be multiplied by \( 1/(1 - \rho^2) \) if we wish to preserve the same standard errors on the individual parameters.

How bad this problem can get is an empirical question about how large \( \rho \) tends to be, and unfortunately in many practical situations, it tends to be large enough to cause significant difficulties. For instance, \cite{bloom2020ideas}'s US TFP measure and their ``number of scientists" R\&D input measure have a correlation coefficient of \( \rho \approx 0.974 \). This makes any attempt at identifying the value of \( \lambda \) and \( \beta \) individually from their time series hopeless; as the sample size of \( \approx 67 \) data points, one per year from 1948 to 2014, is cut down to an effective sample size of only \( \approx 3 \) because of the extreme multicollinearity. This is not an obstacle to obtaining good estimates of \( r = \lambda/\beta \), but it does block estimating the two exponents in the Jones law of motion separately.

\section{Case studies}
\label{sec:case-studies}

In this section, we report three case studies in which we apply the methods from Section \ref{sec:estimation-strategies} to three different domains: the United States TFP data from \cite{bloom2020ideas}, software efficiency estimates for the Stockfish chess engine over time, and other miscellaneous software domains for which we do not have much data. We are specifically interested in software efficiency because we wanted to have estimates of model parameters for software domains for reasons independent of this work. 

We use MLE methods for the first two domains and Bayesian methods for the final domain. Each section is structured to discuss the data we have about the domain, the exact model we fit to the data, and the results we obtain after model fitting, in that order.

\subsection{TFP in the United States}
\label{sec:tfp-united-states}

The data for this case study is taken from \cite{bloom2020ideas}, and we compare our results with the results reported by them whenever possible. The central estimate reported by \cite{bloom2020ideas} for aggregate TFP in the US economy is \( \beta \approx 3.1 \) conditional on assuming \( \lambda = 1 \), corresponding to \( r \approx 1/\beta \approx 0.32 \). Though they use the approach in Appendix \ref{sec:diminishing-returns} to obtain this result, it turns out to be close to the value we get from dividing the growth rates of the outputs and inputs per \hyperref[sec:naive-method]{the naive method}: from 1948 to 2014, US TFP grew at an average rate of \( 1.41 \% \) per year, compared to a \( 5.15 \% \) per year growth rate in research inputs, suggesting \( r \approx 1.41/5.15 \approx 0.27 \).

\begin{figure}[h]
\centering
\includegraphics[scale=0.75]{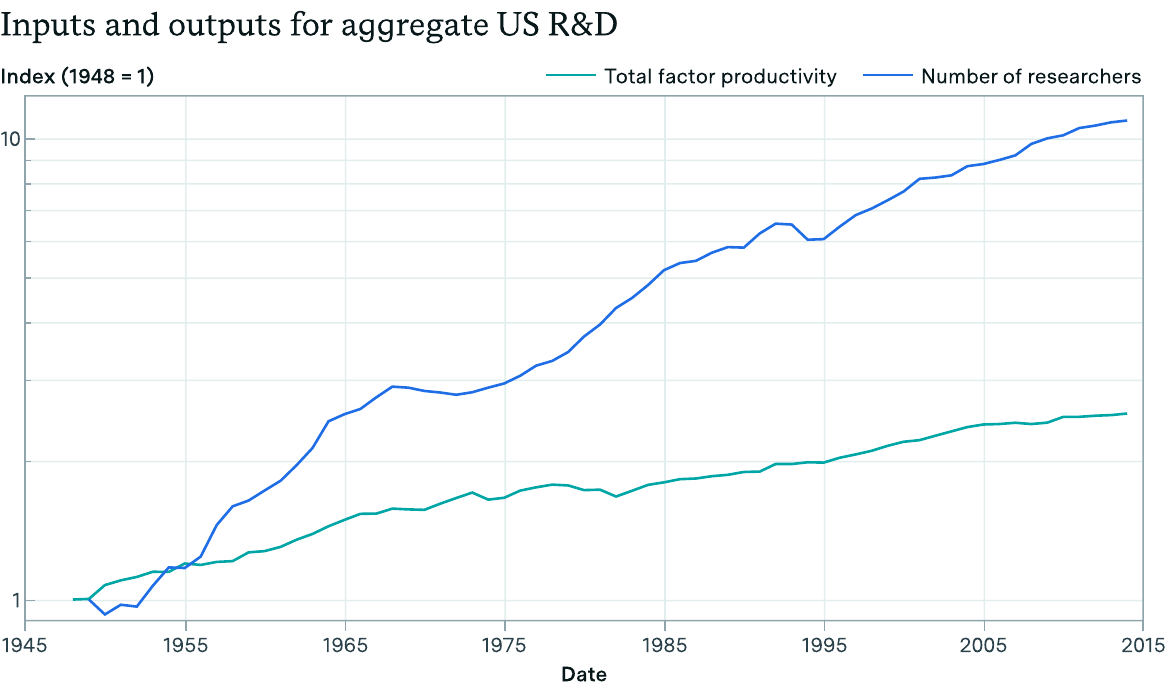}
\caption{Estimates of US total factor productivity and researcher population from \cite{bloom2020ideas}. The values are normalized to an index that is equal to \( 1 \) in the year 1948.}
\label{fig:input-output-tfp}
\end{figure}

There is plenty of data to support the use of the stochastic methods from Section \ref{sec:randomness} as well: we have TFP measurements for every year from 1948 to 2014 inclusive, and corresponding estimates of the number of researchers in the US economy over the same period. This gives us a total of \( n = 2014 - 1948 = 66 \) data points to work with.\footnote{Not adding \( 1 \) to this number is correct because each data point is comprised of a pair \( (A(t_1), A(t_2)) \) along with knowledge of the values of \( I \) from \( t_1 \) to \( t_2 \). \( 67 \) values known for \( A \) are only \( 66 \) useful data points for our purposes.} Figure \ref{fig:input-output-tfp} shows what the data looks like.

For the sake of completeness, we describe here how \cite{bloom2020ideas} obtain their estimates for the number of researchers in the US economy. They take the intellectual property investment time series from \cite{usbea2023ggi} and \cite{usbea2023gpdi}, add them up, then divide this by estimates of the wages of researchers obtained at an annual frequency. They currently proxy for the wages of researchers by looking at mean earnings for males with four or more years of college or graduate school education, with data taken from the Current Population Survey.

This approach already has potential biases, some of which they acknowledge: for instance, a college graduate in 1949 is very different from a college graduate in 2015. It is also not clear that intellectual property expenditures are necessarily the best way to think about the research inputs that go into raising TFP. These are the standard problems from Section \ref{sec:measurement-challenges} that come up routinely when we try to estimate returns to research effort in any domain, so we will not say more about them here.

\begin{figure}[ht]
\centering
\includegraphics[scale=0.75]{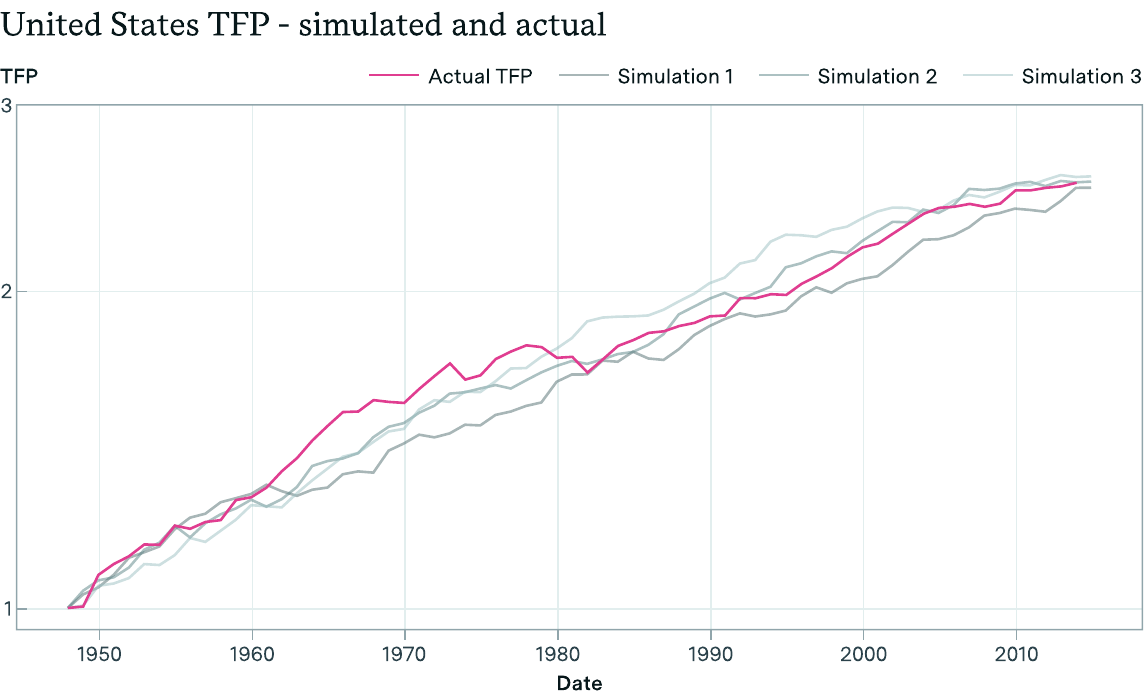}
\caption{Sample simulation runs of the fitted model from Table \ref{tab:third-proposal-tfp-results} compared with the actual TFP data from \cite{bloom2020ideas}. The time series appear quite similar on superficial examination.}
\label{fig:simulations-tfp}
\end{figure}

We now move on to fitting a model to this data. Using the linear regression from Section \ref{sec:linear-regression} is difficult in this case because of the TFP process not being strictly increasing: US TFP decreased between 1976 and 1983, for instance. Consequently, we do not use this method here. As the amount of data we have is sufficient, we instead show the results of using the method outlined in Section \ref{sec:third-proposal}, restricting the Lévy process \( X_p \) to be of drift-diffusion form \( dX_p = \mu \, dt + \sigma \, dW_t \) for \( W_t \) a Wiener process. The results may be found in Table \ref{tab:third-proposal-tfp-results}, and sample simulation runs of the fitted model are presented in Figure \ref{fig:simulations-tfp}.

\begin{table}[ht]
\centering
\small
\begin{tabular}{@{}lccc@{}}
\toprule
                                         & Best fit & Standard error & Standard error \\
                                         &          & (bootstrap)    & (Fisher information matrix) \\ 
\midrule
\( \beta \)                              & 5.42     & 2.51           & 2.55                                       \\ 
\( \lambda \)                            & 1.33     & 0.82           & 0.87                                       \\ 
\( r = \lambda/\beta \)                  & 0.245    & 17.4 ($ \infty $) & N/A                                        \\ 
\( r \) conditional on \( \lambda > 0 \) & 0.251    & 0.056          & N/A                                        \\ 
\bottomrule
\end{tabular}
\caption{The results of fitting the model from Section \ref{sec:third-proposal} to US TFP data from \cite{bloom2020ideas} using maximum likelihood estimation. The Lévy process class was chosen to be the class of drift-diffusion processes. The standard error estimates are obtained in two ways: by bootstrapping the model fit \( n = 100 \) times and by inverting the Fisher information matrix. For \( r \), only the bootstrap method can provide standard errors.}
\label{tab:third-proposal-tfp-results}
\end{table}

\begin{figure}[h]
\centering
\begin{subfigure}[b]{216px}
  \includegraphics[width=\linewidth]{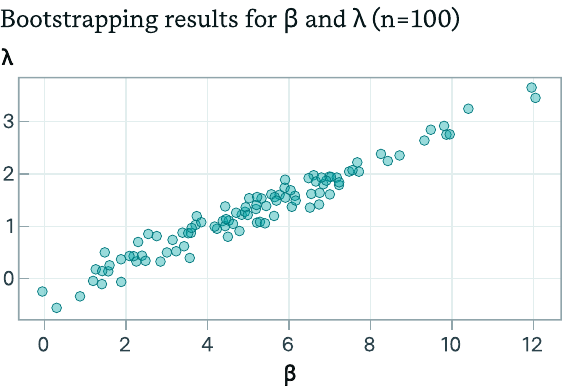}
  \caption{Scatter plot of the values taken by the parameters \( \beta \) and \( \lambda \) in \( n = 100 \) bootstrap runs of the maximum likelihood fit for US TFP.}
  \label{fig:bootstrap-tfp}
\end{subfigure}
\hfill
\begin{subfigure}[b]{219px}
  \includegraphics[width=\linewidth]{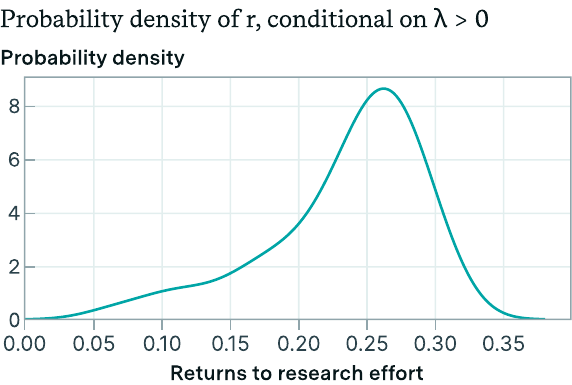}
  \caption{Probability density of \( r \) in the bootstrap, conditional on \( \lambda > 0 \). The smooth-looking plot is generated from the \( 100 \) discrete data points by using a kernel density estimator.}
  \label{fig:returns-density-tfp}
\end{subfigure}
\caption{The bootstrap distribution of model parameters $\lambda$, $\beta$ and returns to research $r$ for US TFP data.}
\end{figure}

The point estimate \( r = 0.245 \) is somewhat lower than the estimate reported in \cite{bloom2020ideas}, but close enough that it is not a cause for concern. However, the standard error explodes: with \( n = 100 \) bootstrap samples it is equal to \( 17.4 \) in our run. This is because the bootstrapping distribution of \( r \) is roughly the ratio of two imperfectly correlated Gaussians \( \lambda \) and \( \beta \), so its standard error should really be infinite. Figure \ref{fig:bootstrap-tfp} is a scatter plot that illustrates this behavior.

Given that the problematic points have \( \lambda < 0 \), which is unrealistic\footnote{As it would imply that progress slows down as the inputs going towards R\&D increase.}, we might wonder what the distribution of \( r \) looks like when we condition on \( \lambda > 0 \) in the bootstrap. In this case, as reported in Table \ref{tab:third-proposal-tfp-results}, the median estimate for the returns is \( 0.251 \), with a much more reasonable standard error of \( 0.056 \). The distribution of \( r \) is skewed with a heavier left tail, as can be seen in Figure \ref{fig:returns-density-tfp}.

As discussed earlier in Section \ref{sec:multicollinearity}, the linear correlation of \( 0.974 \) between the input and output time series in this context makes it impossible to identify \( \beta \) or \( \lambda \) individually with any reasonable degree of confidence. In Table \ref{tab:third-proposal-tfp-results}, this is visible in the high individual standard errors we obtain for \( \lambda \) and \( \beta \) in spite of the large sample size. The small standard errors for \( r \) conditional on \( \lambda > 0 \) show that under an assumption like \( \lambda = 1 \) the standard errors would become more manageable, as expected; but the correlation makes it impossible to jointly identify \( \beta \) and \( \lambda \).

Notably, this means that we cannot reject the hypothesis that \( \lambda = 0 \) based on our standard error estimates, and using a likelihood ratio test on the hypothesis \( \lambda \neq 0 \) only gets a \( p \)-value of \( 0.13 \) if the asymptotic \( \chi^2 \) distribution implied by Wilks' theorem is used.\footnote{See \cite{wilks1938large} for a reference on this result. The test statistic does not follow its asymptotic distribution with our finite sample size, but the \( p \)-value under the asymptotic distribution is nevertheless a useful indicator.} The bootstrap results, in which \( 6 \) out of \( 100 \) points had a negative value of \( \lambda \), suggest \( p \approx 0.06 \) in favor of \( \lambda > 0 \), which also fails to be statistically significant at \( p = 0.05 \) or below.

An appropriate conclusion to draw, therefore, is that the data in \cite{bloom2020ideas} only provides weak evidence that their input measure influences their output measure at all! Unless we already have strong reasons to accept this conclusion on prior belief, the evidence in the paper is simply not strong enough to support it. This casts substantial doubt on the results about US TFP, both in \cite{bloom2020ideas} and also here, as it is not clear we've adequately addressed the input identification problem from Section \ref{sec:unknown-inputs}.

This finding is significant because prior work has criticized \cite{bloom2020ideas}'s approach of concluding that ``ideas are getting harder to find", i.e. \( \beta \gg 0 \), on precisely these grounds. For instance, Section 5 of \cite{guzey_bloom_2021} criticizes the paper for its choice of input measure; arguing that the findings are not robust to reasonable changes to the input measure chosen, especially when it comes to Moore's law (though most of their criticisms extend readily to the TFP case). If the particular input choice chosen by \cite{bloom2020ideas} had achieved a good fit with data, this would be some weak evidence in support of their approach, but here we reach the opposite conclusion that their choice of inputs appears to have no statistically significant connection to TFP. In our view, this gives the criticisms more force than they might have otherwise had.

\subsubsection{Summary}

If we accept that the Jones law of motion as specified in \cite{bloom2020ideas} (as in, with their measure of inputs and outputs) holds in this case with \( \lambda = 1 \), or at least with \( \lambda \gg 0 \), then the estimate of the returns to research effort \( r \) and the qualitative finding that \( \beta \gg 0 \) (ideas get harder to find) from \cite{bloom2020ideas} should both be reliable. However, this is a substantial assumption and is not supported by evidence that is present in the paper or its dataset. If the reader rejects this assumption for TFP, there is little reason for them to trust the precise estimate of \( r \approx 0.32 \), and even dismissing the main thesis \( \beta \gg 0 \) advanced by the paper can be justifiable. A similar criticism applies to the other domains examined in their paper. 

There might, of course, be good prior reasons to suppose that something like this law of motion should hold with \( \lambda \gg 0 \) for something like the input measure used by \cite{bloom2020ideas}. It is straightforward to incorporate such suppositions in the form of priors to the above analysis. However, it is worth making it clear that the raw data do not lend any particular support to the hypothesis that growth in the researcher population has been an important driver of TFP growth in the United States. If we are to believe this, we must believe it for independent reasons.

\subsection{Computer chess}
\label{sec:computer-chess}

\subsubsection{Data description}

As before, we need to collect data on the two time series \( A \) and \( I \) to be able to fit the Jones law of motion. To do this for the Stockfish chess engine, we proxy for \( A \) by combining the Elo estimates from \cite{stockfish2023regression} with the ``Elo from speedups" scaling law reported in \cite{stockfish2023usefuldata}: the software efficiency improvement factor implied by an Elo rating gap of \( \Delta E \) is taken to be \( \exp(\Delta E/C_E) \) for some constant \( C_E = 142.987 \). For \( I \), we use publicly available data on the number of tests completed per day on Fishtest (\cite{stockfish2023fishtest}), the primary distributed testing platform for Stockfish.\footnote{This choice was recommended to us by a major contributor to the Stockfish project.} Overall, this gives us data on \( I \) at a daily frequency, and \( 255 \)\footnote{\( 258 \) if we count three days for which we have the Elo scores of multiple versions reported on a single day.} data points on \( A \).

\begin{figure}[h]
\centering
\begin{subfigure}{0.49\textwidth}
  \includegraphics[width=\linewidth]{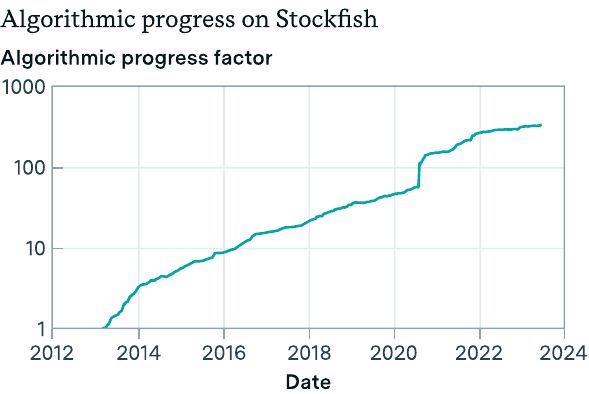}
  \caption{The progress in the algorithmic efficiency of Stockfish over time.}
  \label{fig:stockfish-outputs}
\end{subfigure}
\hfill
\begin{subfigure}{0.49\textwidth}
  \includegraphics[width=\linewidth]{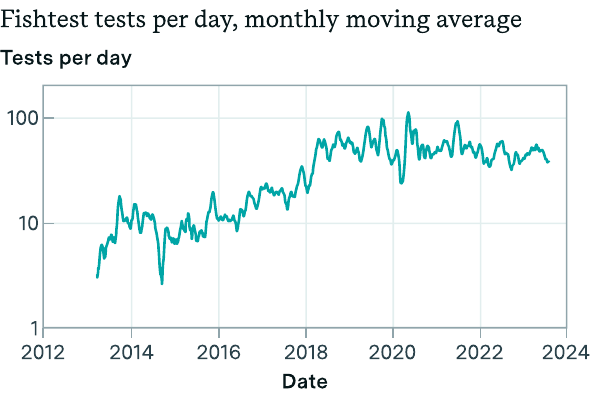}
  \caption{The number of tests completed on Fishtest per day, averaged over the previous 30 days.}
  \label{fig:stockfish-inputs}
\end{subfigure}
\caption{Stockfish Algorithmic Efficiency and Fishtest Tests. The discontinuity in 2020 is notable and was a consequence of the introduction of NNUE, a method of evaluating board positions using a lightweight neural network.}
\end{figure}

Plots of our measures of \( A \) and \( I \) can be found in Figures \ref{fig:stockfish-outputs} and \ref{fig:stockfish-inputs} respectively.

\subsubsection{Results}

\begin{figure}[h]
\centering
\includegraphics[scale=0.75]{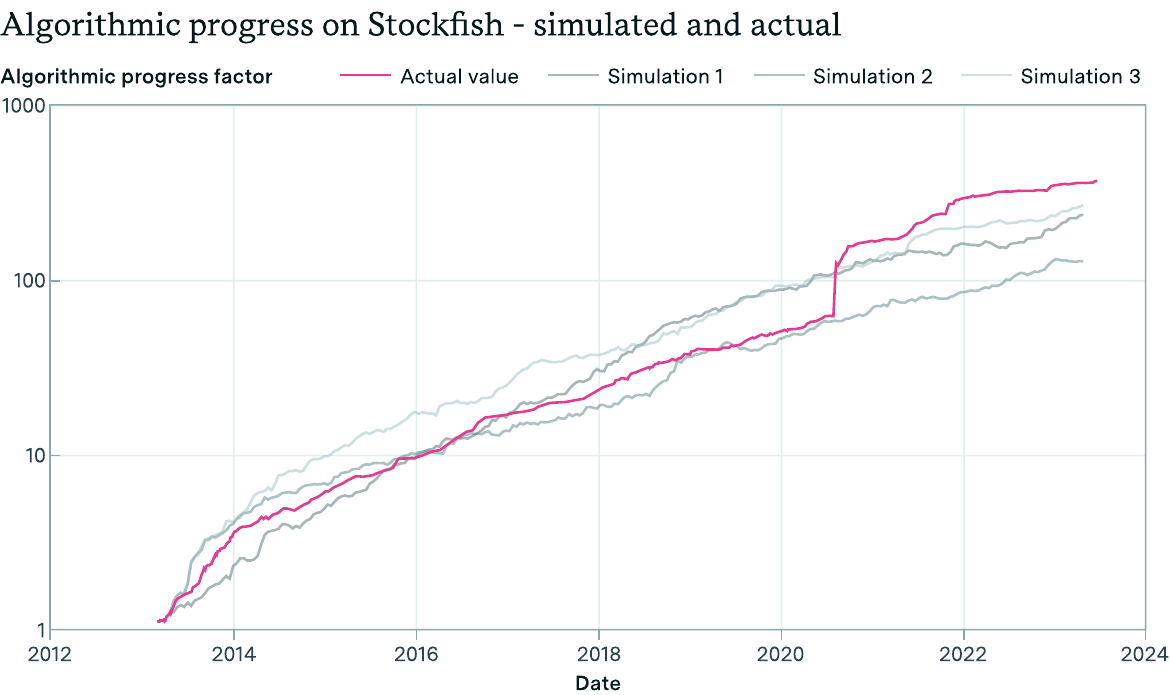}
\caption{Sample simulation runs of the fitted model from Table \ref{tab:third-proposal-chess-results} compared with the actual software progress data obtained from \cite{stockfish2023regression}. Discontinuities of the kind seen in 2020 are rather uncommon even with a stable process taken as \( X_p \).}
\label{fig:simulations-stockfish}
\end{figure}

As before, we fit the model from Section \ref{sec:third-proposal} to the data we have. The only difference in the model from that used in the previous section \ref{sec:tfp-united-states} is that in light of the skewed and discontinuous nature of the software progress time series in Figure \ref{fig:stockfish-outputs}, we broaden the class of Lévy processes to include all stable processes with maximal skewness parameter, i.e. all processes whose increments follow a stable distribution with an almost surely positive jump component. As a Wiener process is stable without any jump component, this includes as a special case all drift-diffusion processes. The results can be found in Table \ref{tab:third-proposal-chess-results}, and sample simulation runs of the fitted model are presented in Figure \ref{fig:simulations-stockfish}.
\begin{table}[h]
\centering
\small
\begin{tabular}{@{}lll@{}}
\toprule
                                         & Best fit & Standard error (bootstrap) \\
\midrule
\( \beta \)                              & 0.476    & 0.066                      \\
\( \lambda \)                            & 0.392    & 0.079                      \\
\( r = \lambda/\beta \)                  & 0.825    & 0.15                       \\
\bottomrule
\end{tabular}
\caption{The results of fitting the model from Section \ref{sec:third-proposal} to Elo and test count data from Stockfish. The Lévy process class was chosen to be the class of stable processes. The standard error estimates are obtained by bootstrapping the model fit \( n = 100 \) times.}
\label{tab:third-proposal-chess-results}
\end{table}

The situation is markedly improved when compared to our estimation with TFP data in Section \ref{sec:tfp-united-states}: we get reasonably small standard errors not just on \( r \) but also on the individual parameters \( \beta \) and \( \lambda \). The basic reason for this is apparent even upon visually inspecting the two plots in Figures \ref{fig:stockfish-inputs} and \ref{fig:stockfish-inputs}: the linear correlation between them is much weaker than it was for US TFP data. If we repeat the same likelihood ratio test on the hypothesis \( \lambda \neq 0 \) that we used in Section \ref{sec:tfp-united-states}, we get a \( p \)-value of \( \approx 0.004 \), which is statistically significant even at a threshold of \( p = 0.01 \). Consequently, in the case of Stockfish, we have some evidence from the data alone that the input measure actually has some influence on the output measure.

To further test the goodness of fit of the model, we run cross-validation by fitting both the exogenous model with \( \lambda = 0 \) and the model where \( \lambda \) is allowed to freely vary to the first \( 80 \% \) of our data points and compare log-likelihoods of both fitted models on the validation set of the remaining \( 20 \% \) data points. The model where \( \lambda \) can freely vary beats the model with the enforced \( \lambda = 0 \) constraint by around \( 2.79 \) nats, confirming the finding from the previous paragraph that the model with freely varying \( \lambda \) is a better model, albeit only with a slight advantage over the purely exogenous model.

It is important to note that we chose the \( 80 \% \) threshold for the cross-validation deliberately to include the NNUE discontinuity in the training set. Otherwise, both models find values of \( \alpha_p \) close to \( 2 \), and goodness of fit on the validation dataset reduces to which fitted model had a smaller value of \( \alpha_p \) to better account for the NNUE discontinuity.

We can also compare the results here with what we would have obtained had we used a more naive method, such as dividing the output and input growth rates as explained in Section \ref{sec:naive-method}. The monthly moving average of the inputs grew at a rate of \( 23 \%/\text{year} \) while the software efficiency of Stockfish grew at a rate of \( 55 \%/\text{year} \), so the naive division would yield \( r = g_A/g_I = 0.55/0.23 \approx 2.4 \). This is almost three times larger than our point estimate of \( r \approx 0.825 \) and is a good example of how the naive method can mislead when the conditions needed for its validity are not satisfied.

\begin{figure}[h]
\centering
\includegraphics[scale=0.75]{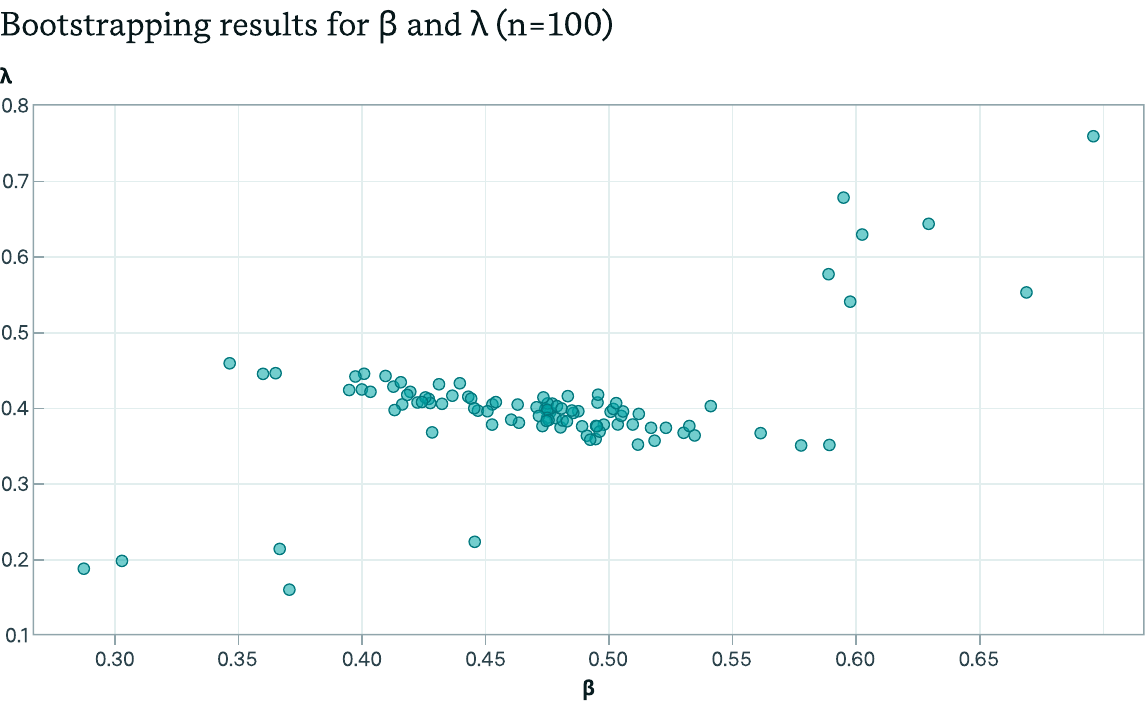}
\caption{Scatter plot of the values taken by the parameters \( \beta \) and \( \lambda \) in \( n = 100 \) bootstrap runs of the maximum likelihood fit for Stockfish.}
\label{fig:bootstrap-stockfish}
\end{figure}

To visualize more information about the behavior of the model, it is worthwhile to inspect a scatter plot of the two dimensionless parameters \( \beta, \lambda \) over different bootstrapping runs. We provide such a plot in Figure \ref{fig:bootstrap-stockfish}.

\subsubsection{Endogeneity problems}

We do not attempt to address potential endogeneity issues when doing this estimation, which could be a factor that overturns our result that input influence on the rate of progress is significant. We can imagine that the level of inputs going into R\&D is itself an endogenous variable influenced by how promising the area of research seems at the moment. In this case, depending on people's preferences, our results here might end up understating or overstating the true sensitivity of the rate of progress to our specific choice of inputs.

The most common way to control for endogeneity in econometrics is to use an instrument: that is, find some variable \( Z \) that we expect to causally influence \( \dot A/A \) only through its influence on \( I \). Finding such an instrument is tricky, however, and we have not managed to find a \( Z \) for which we both have sufficiently abundant data and for which the associated causality assumption seems significantly more plausible than assuming strict exogeneity of the inputs themselves.

These concerns apply just as much to the results from Section \ref{sec:tfp-united-states}, but in that section, we fail to obtain statistical significance even without an instrument. While it is possible that a good choice of instrument would in fact reduce the amount of noise, we consider this fairly unlikely, as instruments typically increase the amount of noise in estimators in exchange for reducing bias. So we believe endogeneity is only a serious problem for this section.

\begin{comment}
The scatter plot in Figure \ref{fig:bootstrap-stockfish} is rather suspicious because of the combination of clustering in the center with outliers far away from the main cluster. We guessed that this was because of the three individual data points in 2020 corresponding to the introduction and optimization of NNUE (efficiently updatable neural network: these data points are correlated with each other, so our model which treats increments in \( A^{\beta} \) as being independent conditional on \( I \) is inaccurate here.
\end{comment}

\subsubsection{Summary}

Software progress in Stockfish is a domain where we have some evidence that the measure of inputs we've chosen has some impact on efficiency. However, while the evidence is statistically significant, it is weaker than we would like (only a few nats in cross-validation, and \( p = 0.004 \) in a likelihood ratio test) which makes it plausible that additional controls or changes in model specification could overturn the finding.

This is also the domain in which the naive estimate of \( r = g_A/g_I \) has its worst performance: as \( r_{\text{MLE}} \approx 0.82 \) and \( r_{\text{naive}} \approx 2.4 \), the naive method overestimates the ``true value" we get from maximum likelihood estimation by a factor of \( \approx 3 \). This shows that while the naive method can be useful to get a quick ballpark estimate of \( r \), it is not a substitute for a more careful analysis of the time series data.

\subsection{Other domains of software R\&D}
\label{sec:other-domains}

Sections \ref{sec:tfp-united-states} and \ref{sec:computer-chess} provided examples of the kinds of results we can get when we have high-frequency data. However, in many domains, we might only have limited information about the output time series \( A \). For instance, a common situation is for software efficiency improvements to be reported over long time periods of one or two decades, and we might not have finer-grained information about \( A \) beyond that. In this case, the Jones law of motion is underidentified, so methods such as MLE will be hopeless. However, if we are willing to use some prior knowledge about what we expect the parameters of a stochastic law of motion to be like, we can use even very limited data for a Bayesian update on the prior per Section \ref{sec:bayesian}.

Our knowledge about the domains of software R\&D we consider in this subsection falls into the underidentified category, and so the approach we choose is Bayesian instead of frequentist. To be conservative, we use fairly uninformative priors for relevant parameters, e.g. half Cauchy with unit scale for dimensionless parameters such as \( \lambda, \beta \) that are restricted to be positive. This helps ensure that our choice of prior does not bias our conclusions in an unduly aggressive manner. Somewhat surprisingly, we discover that updating on only a single data point can substantially narrow our prior over relevant parameters such as the returns to research effort \( r = \lambda/\beta \).

\subsubsection{Data description}

Our data for software efficiency comes from looking at domain-specific papers about software progress in different domains. We consider four domains: computer vision, reinforcement learning, SAT solvers, and linear programming. Table \ref{tab:other-domains-output-data} provides a summary of our efficiency data.

\begin{table}[h]
\centering
\small
\begin{tabular}{@{}llll@{}}
\toprule
                                         & Doubling time                                 & Time period                       & Reference \\
\midrule
Computer vision                          & 9 months                                      & 2012 to 2022                      & \cite{erdil2022algorithmic}                \\
RL sample efficiency                     & 11 months                                     & 2015 to 2019                      & \cite{dorner2021measuring}                 \\
SAT solvers                              & 2 years                                       & 1997 to 2018                      & \cite{fichte2020time}                      \\
Linear programming                       & $ 20/\log_2(9) \approx 6.31 \text{ years} $   & 1998 to 2018                      & \cite{koch2022progress}                    \\
\bottomrule
\end{tabular}
\caption{A summary of our software efficiency data across different domains.}
\label{tab:other-domains-output-data}
\end{table}

As discussed in Section \ref{sec:unknown-outputs}, the fact that software efficiency is properly conceived of as a multidimensional latent variable means that these doubling times are necessarily coarse approximations that are averaged over a suitably large class of problems. For instance, there have been SAT instances that have seen faster progress than a doubling per 2 years, and those that have seen substantially slower progress. The headline figure of 2 years is based on the abstract to \cite{fichte2020time} stating that ``Our findings show that the progress on the algorithmic side has at least as much impact as the progress on the hardware side." combined with the classical Moore's law doubling time of 2 years for hardware efficiency.

For input data, we use the number of unique authors that have published papers recorded in the OpenAlex database about these subjects. Here, the precise definition of ``about" is important: for each category, we define an intersection of OpenAlex concepts that we think captures the papers we care about best. The precise concepts we use can be found in Table \ref{tab:other-domains-input-data}.

Our results are sensitive to which measure of inputs we choose, as discussed in Section \ref{sec:unknown-inputs}, but it is not feasible to make the choice in a way that is more principled because the lack of data prevents adequate model comparison. Conditional on the input choices being good, the results in this section should be useful; and if the input choices are poor, this section should still be a useful demonstration of how to use Bayesian methods to make inferences about the Jones law of motion in the data-scarce regime.

\begin{table}[h]
\centering
\small
\begin{tabular}{@{}ll@{}}
\toprule
                                         & OpenAlex concepts                                                \\
\midrule
Computer vision                          & C31972630 (computer vision) AND C108583219 (deep learning)       \\
RL sample efficiency                     & C97541855 (reinforcement learning)                               \\
SAT solvers                              & C6943359 (boolean satisfiability problem)                        \\
Linear programming                       & C41045048 (linear programming)                                   \\
\bottomrule
\end{tabular}
\caption{The OpenAlex concepts we take intersections of to obtain our input measures for the different software domains we consider.}
\label{tab:other-domains-input-data}
\end{table}

\subsubsection{Results}

We choose to use the model from Section \ref{sec:feller} with \( X_p \) restricted to be a drift-diffusion process, as this process has the most realistic noise structure in the case \( \lambda = 0 \). In this case, as explained in Section \ref{sec:feller}, we are able to leverage the explicit forward probability densities from \cite{roodman2020probability} to provide the log-likelihoods necessary for a Bayesian update.

The choice of priors is also fairly important. Recall Equation \ref{eq:third-proposal-gaussian-law-of-motion-two}:

\begin{equation}
\tag{\ref{eq:third-proposal-gaussian-law-of-motion-two} revisited}
    \frac{dA}{A} = \theta A^{-\beta} I^{\lambda} \, dt + \sigma A^{-\beta/2} I^{\lambda/2} \, dW_t
\end{equation}

Of the four parameters of this SDE, \( \beta, \lambda \) are dimensionless and we pick independent half-Cauchy priors with unit scale over both of them. However, the situation is not as straightforward for the remaining parameters \( \theta, \sigma \), as these two parameters have complicated dimensions. Using square brackets to denote dimensions, we have \( [\theta A^{-\beta} I^{\lambda} \, dt] = 1 \) and so \( [\theta] = [A]^{\beta} [I]^{-\lambda} [t]^{-1} \). Similarly, \( [\sigma] = [\theta]^{1/2} = [A]^{\beta/2} [I]^{-\lambda/2} [t]^{-1/2} \). This is suggestive that we should find some scales \( A_s, I_s, \Delta t_s \) and then pick half Cauchy priors with unit scale over the two dimensionless quantities
\begin{align}
\label{eq:dimensionless-constants}
    \theta_s &= \theta A_s^{-\beta} I_s^{\lambda} (\Delta t_s) \\
    \sigma_s &= \sigma A_s^{-\beta/2} I_s^{\lambda/2} (\Delta t_s)^{1/2}.
\end{align}
There is one caveat: the positivity constraint \( \theta > \sigma^2 (1-\beta)/2 \) discussed in Section \ref{sec:feller}. Enforcing this constraint is equivalent to asking for \( \theta_s > \sigma_s^2 (1-\beta)/2 \), and in order to do this we choose \( \sigma_s \sim \operatorname{HalfCauchy}(1), \, \theta_s \sim \sigma_s^2 \max((1-\beta), 0)/2 + \operatorname{HalfCauchy}(1) \). This prior ensures that all parameter values in the support of our prior are in fact admissible.

We make the choices \( A_s = A_{\text{initial}}, \, I_s = I_{\text{initial}} \): this is nothing more than a change of units that effectively enforces \( A_{\text{initial}} = I_{\text{initial}} = 1 \). The choice of \( \Delta t_s \) is harder. We make this choice in a way that makes joint exponential growth in \( A \) and \( I \) the typical prior outcome, as this matches our observations across many different domains. To do this, we define the average input growth rate \( g_I = \log(I_{\text{final}}/I_{\text{initial}})/(t_{\text{final}} - t_{\text{initial}}) \) and use the naive estimate \( g_A = r g_I = \lambda g_I/\beta \) as a baseline for the initial growth rate. Since the initial average growth rate of \( A \) is given precisely by \( 1/\Delta t_s \), this means we choose \( \Delta t_s = (\lambda g_I/\beta)^{-1} \).

Our final priors are therefore \( \lambda, \beta, \theta_s - \sigma_s^2 \max((1-\beta), 0)/2, \sigma_s \sim \text{HalfCauchy}(1) \) with the priors for all four expressions being jointly independent. It is possible to be more aggressive by choosing a prior that is more strongly bounded away from zero and infinity, but we stick to Cauchy priors to see how big the effect from updating on one data point can be even with relatively uninformative priors. Even these choices turn out to be not as benign as one might hope: the choice of \( \Delta t_s \) in particular is quite significant, as when updating only on a single data point the Bayesian update has a tendency to assume that rates of progress much slower than \( 1/\Delta t_s \) are in part caused by diminishing returns, i.e. small values of \( r \).

\begin{figure}[h]
\centering
\includegraphics[scale=0.75]{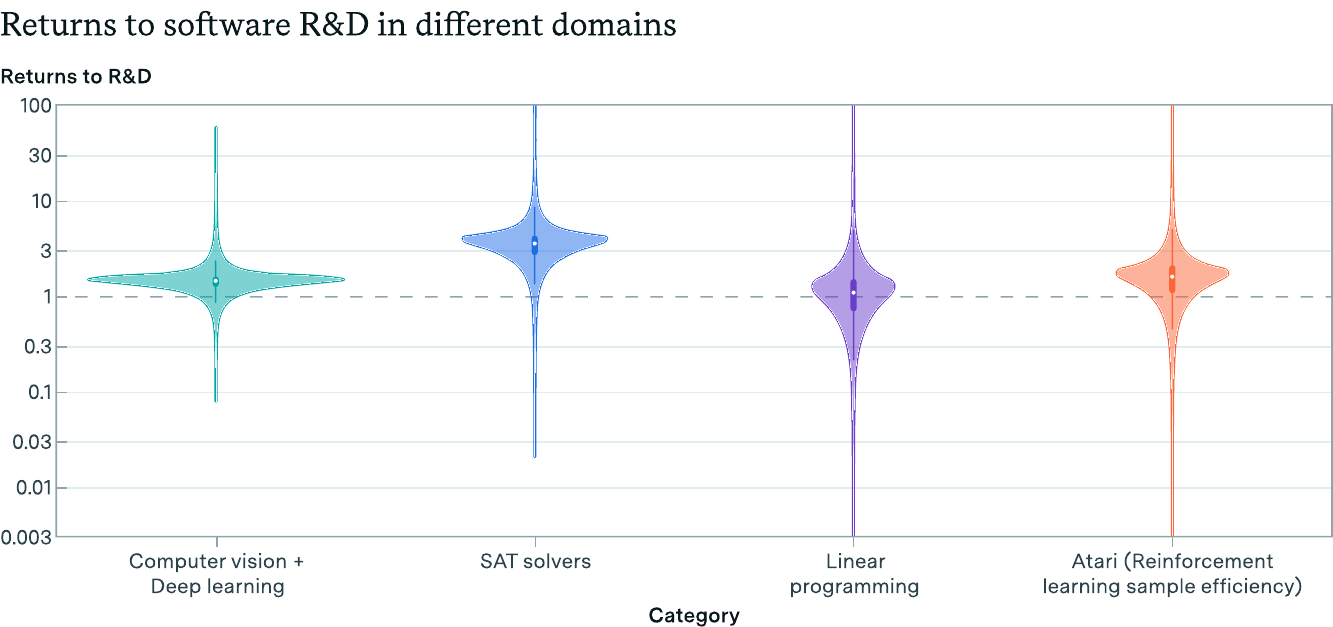}
\caption{Violin plot of the posterior distributions we obtain for the parameter \( r \) across categories. The $r=1$ threshold is marked by the dashed horizontal line for ease of viewing, though it does not always have qualitative significance.}
\label{fig:other-domains-results-returns-violin}
\end{figure}

The results of our Bayesian analysis can be found in Tables \ref{tab:other-domains-results-beta-lambda-table} and \ref{tab:other-domains-results-returns-table}. Figure \ref{fig:other-domains-results-returns-violin} shows the posterior distributions we obtain over the returns to software R\&D parameter \( r = \lambda/\beta \) in a more accessible violin plot format.

\begin{table}[h]
\centering
\small
\begin{tabular}{@{}lll@{}}
\toprule
& \( \beta \) & \( \lambda \) \\
\midrule
Computer vision & 0.985 (0.224 to 4.050) & 1.410 (0.290 to 6.021) \\
RL sample efficiency & 1.023 (0.212 to 3.914) & 1.482 (0.266 to 6.650) \\
SAT solvers & 0.648 (0.139 to 2.891) & 2.143 (0.387 to 11.312) \\
Linear programming & 1.290 (0.254 to 4.953) & 1.259 (0.222 to 5.772) \\
\bottomrule
\end{tabular}
\caption{Estimates of \( \beta \) and \( \lambda \) according to their posterior distributions across categories. We report the median as the point estimate outside parentheses and the central 90\% confidence interval (5th to 95th percentiles) in parentheses.}
\label{tab:other-domains-results-beta-lambda-table}
\end{table}

The results in Table \ref{tab:other-domains-results-beta-lambda-table} may initially be rather hard to interpret, so it is useful to keep what we would obtain from just the prior distribution of half Cauchy with unit scale. The quantile function of the half Cauchy distribution with unit scale is \( Q(\alpha) = \tan(\alpha \pi/2) \), so if we just used the prior for \( \beta, \lambda \) without any Bayesian update, we would expect to get a median of 1 with a 90\% confidence interval of \( \tan(0.05 \cdot \pi/2) \approx 0.078 \) to \( \tan(0.95 \cdot \pi/2) \approx 12.7 \). Looking at the results in Table \ref{tab:other-domains-results-beta-lambda-table}, the Bayesian update narrows the distribution of the parameters relative to this baseline in all cases, though the updates on \( \beta \) are stronger than those on \( \lambda \).

\begin{table}[h]
\centering
\small
\begin{tabular}{@{}llllll|l@{}}
\toprule
& 5th & 25th & \textbf{50th} & 75th & 95th & Naive estimate \\
\midrule
Computer vision & 0.821 & 1.243 & \textbf{1.437} & 1.597 & 2.420 & 1.45 \\
RL sample efficiency & 0.459 & 1.103 & \textbf{1.583} & 2.014 & 3.673 & 1.66 \\
SAT solvers & 1.279 & 2.642 & \textbf{3.542} & 4.230 & 6.897 & 4.17 \\
Linear programming & 0.245 & 0.681 & \textbf{1.077} & 1.508 & 3.095 & 1.51 \\
\bottomrule
\end{tabular}
\caption{Percentiles of the posterior distribution of the returns to software parameter \( r \) across categories. The posterior medians are in bold. The naive estimate column reports the results of estimating \( r \) by dividing the growth rates of the outputs and the inputs as explained in Section \ref{sec:naive-method}.}
\label{tab:other-domains-results-returns-table}
\end{table}

The narrowing of the distribution is even more pronounced in Table \ref{tab:other-domains-results-returns-table}. This is because the prior for \( r \) is the ratio of two independent half-Cauchy random variables with unit scale and because the half-Cauchy distribution is invariant under taking reciprocals, so the prior for \( r \) is the distribution of the product of two independent half-Cauchy random variables with unit scale, which is identical to the distribution of \( |XY| \) where the variables \( X, Y \) are independent and Cauchy with unit scale. This distribution can be inferred from the results in \cite{springer1966distribution} and has density
\begin{equation} f_{\text{half cauchy product}}(x) = \frac{4}{\pi^2} \frac{\log x}{x^2 - 1} \end{equation}
on the positive real numbers. Numerical integration then gives us that the prior distribution for \( r \) has a median of \( 1 \) (this is obvious from the \( x \to 1/x \) symmetry) and a 90\% confidence interval of  \( 0.0267 \) to \( 37.5 \). This interval is roughly three orders of magnitude wide, and in all cases, we observe a substantial reduction of this prior uncertainty after performing a Bayesian update on a single data point.

\subsubsection{Summary}

The Bayesian method outlined in Section \ref{sec:bayesian} is useful for getting tentative estimates of model parameters even in the data-scarce regime, and updating even on a single observation can substantially reduce prior uncertainty. However, it should be used with caution, because data scarcity makes it impossible to test the model's goodness-of-fit with data. If the model itself is wrong or misspecified, the conclusions of the Bayesian method are of little value as they are by definition conditional on the model class being correctly chosen.

Despite its shortcomings in the data-scarce regime, the Bayesian method should still be favored over the naive estimate of \( r = g_A/g_I \), as the naive method shares the same model validity problems as the Bayesian method and has additional problems on top of that. This can be seen in Table \ref{tab:other-domains-results-returns-table} and Section \ref{sec:computer-chess}: the naive method often gives answers that are quite inaccurate when compared with more reliable, likelihood-based methods.

\section{Conclusion}

This paper has reviewed strategies for estimating key parameters governing the dynamics of idea production, a problem of fundamental importance to innovation theory and the study of economic growth. Through case studies and examples, we demonstrated the application of these strategies across diverse domains, ranging from aggregate productivity measurement to software R\&D.

We highlight a few key obstacles and illustrate these with the help of case studies:
\begin{itemize}
    \item Identifying appropriate input and output measures is extremely challenging (Sections \ref{sec:unknown-inputs}, \ref{sec:unknown-outputs}). Using incorrect measures undermines all subsequent analysis.
    \item For U.S. TFP, the statistical evidence that the chosen research input measure substantively impacts the output metric is remarkably weak (Section \ref{sec:tfp-united-states}). A likelihood ratio test fails to reject at any standard significance level the null hypothesis that the input measure has precisely zero influence on the output. This absence of empirical validation gives no support to the core modeling assumption that growth in the researcher population drove measured TFP gains. Unless compelling independent reasons exist to maintain this questionable assumption, the ensuing conclusions regarding the extent of diminishing returns and precise quantitative estimates of the returns to research effort are on shaky grounds.
    \item Evidence of input influence is statistically significant (\( p=0.004 \)) for computer chess. Though statistically significant, the analysis provides only weak evidence of input influence in computer chess because we do not have any means of correcting for endogeneity.
    \item Naive methods such as dividing the output and input growth rates are a decent default baseline for back-of-the-envelope calculations but can diverge from more rigorous estimates by a factor of \( 3 \) or more in realistic scenarios. For instance, dividing the growth rates overestimates returns likely substantially for computer chess in Section \ref{sec:computer-chess}. Maximum likelihood and Bayesian techniques are relatively better, but remain sensitive to misspecification of the underlying model.
    \item With scarce data, model validity is hard to assess. Even assuming the validity of the model, parameter estimates remain sensitive to exactly how priors are specified (Section \ref{sec:other-domains}). So inferences made in this regime are tentative, and even Bayesian posterior confidence intervals do not fully capture the uncertainty we should have over relevant parameters conditional on only a single data point.
\end{itemize}
While the advanced statistical techniques synthesized in this paper can certainly aid the estimation process, they ultimately cannot resolve these core problems. Careful empirical work accounting for these issues via cross-validation and other techniques, paired with domain expertise to improve model specification, remains essential to making credible progress on this estimation problem.

We further present valuable object-level results for software R\&D spanning computer chess (Stockfish chess engine), computer vision, reinforcement learning, SAT solvers, and linear programming. In the domain with the best data, computer chess, we find a estimate of the returns to research of $0.825$. This is found to be statistically significantly greater than $0$ but not less than $1$ (the cutoff for hyperbolic growth in such endogenous growth models, see \ref{sec:importance-of-r}). The 90\% posterior confidence intervals for returns are: computer vision $[0.821, 2.420]$, reinforcement learning sample efficiency $[0.459, 3.673]$, SAT solvers $[1.279, 6.897]$, and linear programming $[0.245, 3.095]$. While median estimates are above $1$, we stress that model validity is hard to assess given the limited data, and conclusions drawn are tentative.

\appendix
\section*{\Large Appendices}

\section{Asymptotics of growth and the returns to scale}
\label{sec:importance-of-r}
In this section we provide an example to illustrate the importance \( r \) typically has in endogenous growth theory models. Suppose that \( A \) is TFP in an \( AK \) growth model, described by the laws of motion:
\begin{align}
    Y &= AK^{\alpha} \\
    \frac{dK}{dt} &\propto Y \\
    \frac{1}{A} \frac{dA}{dt} &\propto A^{-\beta} K^{\lambda},
\end{align}
where the symbol \( \propto \) denotes proportionality, allowing us to omit inessential constants such as saving rates \textit{et cetera}. Denoting the growth rates of \( A, K, Y \) by \( g_A, g_K, g_Y \) respectively, in a steady-state with exponential growth we have:
\begin{align}
    g_Y &= g_A + \alpha g_K \\
    g_K &= g_Y \\
    \lambda g_K &= \beta g_A.
\end{align}
For this system to admit a solution with all growth rates nonzero, we must have that \( \lambda/\beta + \alpha = r + \alpha = 1 \). Moreover, it is also possible to show that the system shows hyperbolic growth and diverges in finite time when \( r + \alpha > 1 \), and it shows polynomial growth when \( r + \alpha < 1 \). Crucially, it is the parameter \( r \) that is important, not \( \lambda, \, \beta \) individually.

\section{Derivation: Dividing the growth rates}
\label{sec:derivation-naive-method}

In this section we lay out the details of results in Section \ref{sec:naive-method} for the naive method. Recall from our discussion that we have
\begin{equation}
\label{eq:ratio-solution}
    \left( \frac{(A(t_2)/A(t_0))^{\beta} - 1}{(A(t_1)/A(t_0))^{\beta} - 1} \right)^{1/\lambda} = \frac{I_{\lambda}(t_0, t_2)}{I_{\lambda}(t_0, t_1)}.
\end{equation}
Now, as \( I_{\lambda}(t_a, t_b) \) is the \( L^{\lambda} \) norm of \( I \) over the time interval \( [t_a, t_b] \), as \( \lambda \to \infty \) it converges to the \( L^{\infty} \) norm, which is equal to the essential supremum of \( I \) over the associated interval.\footnote{This approximation will be good even for small values of \( \lambda \) if the time intervals are chosen to be narrow, but in cases where the law of motion is not exact this increases the amount of noise, so it is preferable to pick the time intervals to be sufficiently wide to get reliable answers from this method in practical situations.}

In particular, if \( I \) is continuous and increasing, we will have
\begin{equation} \lim_{\lambda \to \infty} \frac{I_{\lambda}(t_0, t_2)}{I_{\lambda}(t_0, t_1)} = \frac{I(t_2)}{I(t_1)} .\end{equation}
It follows from this that if \( A(t_2) > A(t_1) > A(t_0) \), the value of \( \beta \) satisfying the equation must likewise diverge to positive infinity; as the mapping
\begin{equation} 
\beta \mapsto \frac{(A(t_2)/A(t_0))^{\beta} - 1}{(A(t_1)/A(t_0))^{\beta} - 1} 
\end{equation}
defines a continuous, strictly increasing function of \( \beta \in \mathbb R \) which is bounded as \( \beta \to -\infty \). The apparent singularity at \( \beta = 0 \) is removable, as the limit of the expression as \( \beta \to 0 \) is well-defined and finite. Viewing the value of \( \beta \) that solves the equation as a function \( \beta(\lambda) \) of \( \lambda\), we then know that the limit
\begin{equation} 
\lim_{\lambda \to \infty} \left( \frac{(A(t_2)/A(t_0))^{\beta(\lambda)} - 1}{(A(t_1)/A(t_0))^{\beta(\lambda)} - 1} \right)^{1/\lambda} = \frac{I(t_2)}{I(t_1)}. 
\end{equation}
Taking logarithms on both sides and using the continuity of the logarithm gives
\begin{equation} 
\lim_{\lambda \to \infty} \frac{\beta(\lambda) \log \left( \frac{A(t_2)}{A(t_1)} \right) + \log \left( 1 - \left( \frac{A(t_2)}{A(t_0)} \right)^{-\beta(\lambda)}  \right) - \log \left( 1 - \left( \frac{A(t_1)}{A(t_0)} \right)^{-\beta(\lambda)} \right)}{\lambda} = \log \left( \frac{I(t_2)}{I(t_1)} \right). 
\end{equation} 
The second and third terms in the numerator are \textit{bounded}, and so as \( \lambda \to \infty \) their ratio with \( \lambda \) tends to zero. It follows that
\begin{equation} \lim_{\lambda \to \infty} \frac{\beta(\lambda) \log \left( \frac{A(t_2)}{A(t_1)} \right)}{\lambda} = \log \left( \frac{I(t_2)}{I(t_1)} \right). \end{equation} 
This is equivalent to
\begin{equation}
% \label{eq:asymptotic-naive-method}
    \lim_{\lambda \to \infty} r(\lambda) = \lim_{\lambda \to \infty} \frac{\lambda}{\beta(\lambda)} = \frac{\log \left( \frac{A(t_2)}{A(t_1)} \right)}{\log \left( \frac{I(t_2)}{I(t_1)} \right)},
\end{equation}
which is what we sought out to show.

\section{Details on approximate diminishing returns estimation}
\label{sec:diminishing-returns}

% \subsection{Approximate estimation based on the diminishing returns}
% \label{sec:diminishing-returns}

An important variant of the technique described in Section \ref{sec:naive-method} is used in \cite{bloom2020ideas} to estimate \( \beta \) when \( \lambda \) is already known. We do not think it can be recommended as it is prone to error, despite being a good approximation in many situations. We report it only for the sake of addressing prior methods used in the literature: if you have enough data to use this method, there are better choices available to you (as discussed in Section \ref{sec:estimation-strategies}).

This particular method is based on the following approach: start with the Jones law of motion
\begin{equation} 
\frac{\dot A}{A} = \theta A^{-\beta} I^{\lambda} \end{equation}
and evaluate it at two times \( t = t_1, t_2 \). Divide through to obtain
\begin{equation} 
\frac{(\dot A/A)_{t=t_2}}{(\dot A/A)_{t=t_1}} = \left( \frac{A(t_2)}{A(t_1)} \right)^{-\beta} \left( \frac{I(t_2)}{I(t_1)} \right)^{\lambda}.
\end{equation}
Taking logarithms on both sides and isolating \( \beta \) gives the equality
\begin{equation} 
\beta = \frac{\log((\dot A/A)_{t=t_1}/I(t_1)^{\lambda}) - \log((\dot A/A)_{t=t_2}/I(t_2)^{\lambda})}{\log(A(t_2)) - \log(A(t_1))} 
\end{equation}
This identity measures \( \beta \) by looking at how much research productivity has fallen over the time interval \( t_1 \) to \( t_2 \), and comparing this with the increase in efficiency over the same period. However, as stated it is not useful because it relies on us needing to know time-intensive quantities such as \( \dot A \) and \( I \), which are susceptible to large amounts of noise in practice. If the differential equation has any noise at all, this estimate is going to be unreliable.

For this identity to be useful, we want to relax it into some time integrated form such as

\begin{equation}
\label{eq:conjecture-dim-returns}
\beta \stackrel{?}{\approx} \frac{\log(\Delta_{t_1, t_2} \log(A)/I_{\lambda}(t_1, t_2)^{\lambda}) - \log(\Delta_{t_3, t_4} \log(A)/I_{\lambda}(t_3, t_4)^{\lambda})}{\log(A(t_3)) - \log(A(t_1))}
\end{equation}

which is what \cite{bloom2020ideas} do. The question mark above the approximate equality sign denotes our uncertainty in whether such an approximation in fact holds or not.

It turns out that this approximation often holds, and in the particular cases where \cite{bloom2020ideas} use it, the results of using it are good. However, it is not guaranteed that the approximation should hold, and there are better methods to use, such as directly using Equation \ref{eq:ratio-identity} to estimate \( \beta \). There is no reason to make the approximation required for this method to hold, as it can often be entirely avoided. For this reason, \textbf{we recommend never using this method}.
% \footnote{The interested reader can find the conditions under which this approximation can expected to be good \hyperref[sec:diminishing-returns]{in the appendices}.}

% In this appendix, we work out the conditions under which the approximate method outlined in Section \ref{sec:diminishing-returns} can be expected to work well. For complete context, this appendix should be read after the reader has finished reading Section \ref{sec:diminishing-returns}.
\subsection{Conditions of applicability}
We now work out the conditions under which this approximate method can be expected to work well. We first express the key identity Equation \ref{eq:solution} as
\begin{equation}
\label{eq:solution-jones-form}
    Q_{\beta}(t_1, t_2) = \frac{\left( \frac{A(t_2)}{A(t_1)} \right)^{\beta} - 1}{\beta} = \theta A(t_1)^{-\beta} I_\lambda(t_1, t_2)^{\lambda}
\end{equation}
where the first equality simply defines the expression \( Q_{\beta}(t_1, t_2) \). This relation is now analogous to the instantaneous Jones law of motion, so using the same manipulations on it gives
\begin{equation}
\label{eq:true-dim-returns}
\beta = \frac{\log(Q_{\beta}(t_1, t_2)/I_{\lambda}(t_1, t_2)^{\lambda}) - \log(Q_{\beta}(t_3, t_4)/I_{\lambda}(t_3, t_4)^{\lambda})}{\log(A(t_3)) - \log(A(t_1))}.
\end{equation}
The problem with this is that it is an implicit relation: \( \beta \) occurs both on the left and the right-hand side, because \( Q_{\beta} \) is a function of \( \beta \). Nevertheless, there is a connection between Equations \ref{eq:conjecture-dim-returns} and \ref{eq:true-dim-returns}. Indeed, if we express
\begin{equation} 
Q_{\beta}(t_1, t_2) = \frac{\left( \frac{A(t_2)}{A(t_1)} \right)^{\beta} - 1}{\beta} = \frac{e^{\beta a(t_1, t_2)} - 1}{\beta} 
\end{equation}

where we adopt the notation \( a(t_1, t_2) = \log(A(t_2)/A(t_1)) \) for convenience, and Taylor expand the exponential in the numerator, we obtain
\begin{equation} 
Q_{\beta}(t_1, t_2) = \frac{\beta a(t_1, t_2) + (\beta^2 a(t_1, t_2)^2)/2 + O(\beta^3 a(t_1, t_2)^3)}{\beta} = a(t_1, t_2) + \frac{\beta a(t_1, t_2)^2}{2} + O(\beta^2 a(t_1, t_2)^3). 
\end{equation}
From this, we deduce the expansion
\begin{equation} 
\log Q_{\beta}(t_1, t_2) = \log a(t_1, t_2) + \frac{\beta a(t_1, t_2)}{2} + O(\beta^2 a(t_1, t_2)^2) 
\end{equation}
which, when substituted into Equation \ref{eq:true-dim-returns}, yields
\begin{align}
\label{eq:connection-dim-returns}
\beta &= \beta_{\text{approx}} + \frac{\beta}{2} \cdot \frac{a(t_1, t_2) - a(t_3, t_4)}{a(t_1, t_3)} + O \left( \beta^2 \cdot \frac{a(t_1, t_2)^2 + a(t_3, t_4)^2}{a(t_1, t_3)} \right) \\
\beta &\approx \frac{\beta_{\text{approx}}}{1 - \frac{a(t_1, t_2) - a(t_3, t_4)}{2 a(t_1, t_3)}}
\end{align}

to top order, where \( \beta_{\text{approx}} \) is \textit{defined} as the value we would estimate if we directly used Equation \ref{eq:conjecture-dim-returns}, i.e. by the expression

\begin{equation} 
\beta_{\text{approx}} = \frac{\log(\Delta_{t_1, t_2} \log(A)/I_{\lambda}(t_1, t_2)^{\lambda}) - \log(\Delta_{t_3, t_4} \log(A)/I_{\lambda}(t_3, t_4)^{\lambda})}{\log(A(t_3)) - \log(A(t_1))}. 
\end{equation}

Equation \ref{eq:connection-dim-returns} tells us how much bias we should expect from using Equation \ref{eq:conjecture-dim-returns} instead of Equation \ref{eq:true-dim-returns} to estimate the value of \( \beta \). The bias scales with \( \frac{a(t_1, t_2) - a(t_3, t_4)}{2 a(t_1, t_3)} \), at least when dropping the higher order terms in the Taylor approximation to \( \log Q_{\beta} \) is valid, so for minimal bias we want to make the time periods \( t_1, t_2 \) and \( t_3, t_4 \) reasonably short while making the time period \( t_1, t_3 \) quite long. 

When \cite{bloom2020ideas} use this for US TFP data, the sampling periods \( t_2 - t_1, \, t_4 - t_3 \) are a decade long, while the interval \( t_3 - t_1 \) is on the order of a century. This means we end up with a bias that is on the order of a few percent in their computation of \( \beta \). However, even this first-order approximation can mislead about the perils of using this method, as the higher order contributions to \( \log Q_{\beta} \) can become large when we are forced to make sampling periods longer to cope with noise, for example.

\printbibliography

@article{erdil2022algorithmic,
  title={Algorithmic progress in computer vision},
  author={Erdil, Ege and Besiroglu, Tamay},
  journal={arXiv preprint arXiv:2212.05153},
  year={2022}
}

@book{arrow1972economic,
  title={Economic welfare and the allocation of resources for invention},
  author={Arrow, Kenneth Joseph},
  year={1972},
  publisher={Springer}
}

@inproceedings{fichte2020time,
  title={A time leap challenge for SAT-solving},
  author={Fichte, Johannes K and Hecher, Markus and Szeider, Stefan},
  booktitle={Principles and Practice of Constraint Programming: 26th International Conference, CP 2020, Louvain-la-Neuve, Belgium, September 7--11, 2020, Proceedings},
  pages={267--285},
  year={2020},
  organization={Springer}
}

@article{bloom2020ideas,
  title={Are ideas getting harder to find?},
  author={Bloom, Nicholas and Jones, Charles I and Van Reenen, John and Webb, Michael},
  journal={American Economic Review},
  volume={110},
  number={4},
  pages={1104--1144},
  year={2020},
  publisher={American Economic Association 2014 Broadway, Suite 305, Nashville, TN 37203}
}

@article{koch2022progress,
  title={Progress in mathematical programming solvers from 2001 to 2020},
  author={Koch, Thorsten and Berthold, Timo and Pedersen, Jaap and Vanaret, Charlie},
  journal={EURO Journal on Computational Optimization},
  volume={10},
  pages={100031},
  year={2022},
  publisher={Elsevier}
}

@article{dorner2021measuring,
  title={Measuring progress in deep reinforcement learning sample efficiency},
  author={Dorner, Florian E},
  journal={arXiv preprint arXiv:2102.04881},
  year={2021}
}

@article{jones1995r,
  title={R \& D-based models of economic growth},
  author={Jones, Charles I},
  journal={Journal of political Economy},
  volume={103},
  number={4},
  pages={759--784},
  year={1995},
  publisher={The University of Chicago Press}
}

@article{romer1990endogenous,
  title={Endogenous technological change},
  author={Romer, Paul M},
  journal={Journal of political Economy},
  volume={98},
  number={5, Part 2},
  pages={S71--S102},
  year={1990},
  publisher={The University of Chicago Press}
}

@article{guzey_bloom_2021,
	title = {Issues with {Bloom} et al's "{Are} {Ideas} {Getting} {Harder} to {Find}?" and why total factor productivity should never be used as a measure of innovation},
	url = {https://guzey.com/economics/bloom/},
	journal = {Guzey.com},
	author = {Guzey, Alexey and Rischel, Eigil and {Applied Divinity Studies} and Anonymous and Anonymous},
	month = sep,
	year = {2021},
}

@misc{ekerdt2023,
        title = {Self-Selection and the Diminishing Returns of Research},
        url = {https://drive.google.com/file/d/1qPdzXu6wLclnHE-YX1eLfRQtkUzCVvD2/view?usp=sharing},
        author = {Ekerdt, Lorenz K.F. and Wu, Kai-Jie},
        month = apr,
        year = {2023}
}

@misc{erdil2023china,
        title = {Is Chinese total factor productivity lower today than it was in 1956?},
        url = {https://www.lesswrong.com/posts/4SsoZLYk6efXFWzY5/is-chinese-total-factor-productivity-lower-today-than-it-was},
        author = {Erdil, Ege},
        month = aug,
        year = {2023}
}

@techreport{davidson2023compute,
        title = {What a compute-centric framework says about takeoff speeds},
        url = {https://www.openphilanthropy.org/research/what-a-compute-centric-framework-says-about-takeoff-speeds/},
        author = {Davidson, Tom},
        institution = {Open Philanthropy},
        year = {2023}
}

@article{barro2013new,
  title={A new data set of educational attainment in the world, 1950--2010},
  author={Barro, Robert J and Lee, Jong Wha},
  journal={Journal of development economics},
  volume={104},
  pages={184--198},
  year={2013},
  publisher={Elsevier}
}

@article{feenstra2015next,
  title={The next generation of the Penn World Table},
  author={Feenstra, Robert C and Inklaar, Robert and Timmer, Marcel P},
  journal={American economic review},
  volume={105},
  number={10},
  pages={3150--3182},
  year={2015},
  publisher={American Economic Association 2014 Broadway, Suite 305, Nashville, TN 37203}
}

@article{jones2022past,
  title={The past and future of economic growth: A semi-endogenous perspective},
  author={Jones, Charles I},
  journal={Annual Review of Economics},
  volume={14},
  pages={125--152},
  year={2022},
  publisher={Annual Reviews}
}

@article{bosworth2008accounting,
  title={Accounting for growth: comparing China and India},
  author={Bosworth, Barry and Collins, Susan M},
  journal={Journal of Economic Perspectives},
  volume={22},
  number={1},
  pages={45--66},
  year={2008},
  publisher={American Economic Association}
}

@article{hernandez2020measuring,
  title={Measuring the algorithmic efficiency of neural networks},
  author={Hernandez, Danny and Brown, Tom B},
  journal={arXiv preprint arXiv:2005.04305},
  year={2020}
}

@article{borak2005stable,
  title={Stable distributions},
  author={Borak, Szymon and H{\"a}rdle, Wolfgang and Weron, Rafal},
  journal={Statistical tools for finance and insurance},
  volume={4},
  year={2005},
  publisher={Springer Berlin},
  url={https://edoc.hu-berlin.de/bitstream/handle/18452/4526/8.pdf}
}

@incollection{cox2005theory,
  title={A theory of the term structure of interest rates},
  author={Cox, John C and Ingersoll Jr, Jonathan E and Ross, Stephen A},
  booktitle={Theory of valuation},
  pages={129--164},
  year={2005},
  publisher={World Scientific}
}

@misc{roodman2020probability,
  title={On the probability distribution of long-term changes in the growth rate of the global economy: An outside view},
  author={Roodman, David},
  journal={Report},
  year={2020},
  publisher={Open Philanthropy San Francisco},
  url={https://www.openphilanthropy.org/wp-content/uploads/Modeling-the-human-trajectory-2.pdf}
}

@article{wilks1938large,
  title={The large-sample distribution of the likelihood ratio for testing composite hypotheses},
  author={Wilks, Samuel S},
  journal={The annals of mathematical statistics},
  volume={9},
  number={1},
  pages={60--62},
  year={1938},
  publisher={JSTOR}
}

@misc{usbea2023gpdi,
    author = {{U.S. Bureau of Economic Analysis}},
    title = {Gross Private Domestic Investment: Fixed Investment: Nonresidential: Intellectual Property Products [Y001RC1A027NBEA]},
    howpublished = {FRED, Federal Reserve Bank of St. Louis},
    year = {2023},
    note = {Retrieved September 12, 2023},
    url = {https://fred.stlouisfed.org/series/Y001RC1A027NBEA}
}

@misc{usbea2023ggi,
    author = {{U.S. Bureau of Economic Analysis}},
    title = {Government Gross Investment: Intellectual Property Products [Y055RC1A027NBEA]},
    howpublished = {FRED, Federal Reserve Bank of St. Louis},
    year = {2023},
    note = {Retrieved September 12, 2023},
    url = {https://fred.stlouisfed.org/series/Y055RC1A027NBEA}
}

@misc{stockfish2023regression,
  author = {{Stockfish}},
  title = {Regression Tests},
  year = {2023},
  url = {https://github.com/official-stockfish/Stockfish/wiki/Regression-Tests},
  note = {Accessed: July 2023}
}

@misc{stockfish2023usefuldata,
  author = {{Stockfish}},
  title = {Useful data},
  year = {2023},
  url = {https://github.com/official-stockfish/Stockfish/wiki/Useful-data##elo-from-speedups},
  note = {Accessed: July 2023}
}

@misc{stockfish2023fishtest,
  author = {{Stockfish}},
  title = {Stockfish Testing Framework},
  year = {2023},
  url = {https://tests.stockfishchess.org/tests},
  note = {Accessed: July 2023}
}

@article{abrilpla2023pymc,
    author = {Abril-Pla, O. and Andreani, V. and Carroll, C. and Dong, L. and Fonnesbeck, C.J. and Kochurov, M. and Kumar, R. and Lao, J. and Luhmann, C.C. and Martin, O.A. and Osthege, M. and Vieira, R. and Wiecki, T. and Zinkov, R.},
    year = {2023},
    title = {PyMC: a modern, and comprehensive probabilistic programming framework in Python},
    journal = {PeerJ Computer Science},
    volume = {9},
    pages = {e1516},
    doi = {https://doi.org/10.7717/peerj-cs.1516}
}

@article{hoffman2014no,
  title={The No-U-Turn sampler: adaptively setting path lengths in Hamiltonian Monte Carlo.},
  author={Hoffman, Matthew D and Gelman, Andrew and others},
  journal={J. Mach. Learn. Res.},
  volume={15},
  number={1},
  pages={1593--1623},
  year={2014}
}

@article{springer1966distribution,
  title={The distribution of products of independent random variables},
  author={Springer, Melvin Dale and Thompson, WE},
  journal={SIAM Journal on Applied Mathematics},
  volume={14},
  number={3},
  pages={511--526},
  year={1966},
  publisher={SIAM}
}

@book{nolan2020univariate,
  title={Univariate stable distributions},
  author={Nolan, John P},
  year={2020},
  publisher={Springer}
}

@article{Pessoa2005IdeasDG,
  title={“Ideas” driven growth: the OECD evidence},
  author={Argentino Pessoa},
  journal={Portuguese Economic Journal},
  year={2005},
  volume={4},
  pages={46-67},
  url={https://api.semanticscholar.org/CorpusID:51999283}
}

@article{Sequeira2020SteppingOT,
  title={Stepping on toes in the production of knowledge: a meta-regression analysis},
  author={Tiago Neves Sequeira and Pedro Cunha Neves},
  journal={Applied Economics},
  year={2020},
  volume={52},
  pages={260 - 274},
  url={https://api.semanticscholar.org/CorpusID:159262716}
}

@article{Neves2018SpilloversIT,
  title={Spillovers in the production of knowledge: A meta-regression analysis},
  author={Pedro Cunha Neves and Tiago Neves Sequeira},
  journal={Research Policy},
  year={2018},
  volume={47},
  pages={750-767},
  url={https://api.semanticscholar.org/CorpusID:85503718}
}

@article{Hall2009MeasuringTR,
  title={Measuring the Returns to R\&D},
  author={Bronwyn H Hall and Jacques Mairesse and Pierre Mohnen},
  journal={Environment for Innovation eJournal},
  year={2009},
  url={https://api.semanticscholar.org/CorpusID:18036207}
}

@article{Herzer2020SemiendogenousVS,
  title={Semi-endogenous Versus Schumpeterian Growth Models: A Critical Review of the Literature and New Evidence},
  author={Dierk Herzer},
  journal={Review of Economics},
  year={2020},
  volume={73},
  pages={1 - 55},
  url={https://api.semanticscholar.org/CorpusID:213425526}
}

@article{Hooker2020TheHL,
  title={The hardware lottery},
  author={Sara Hooker},
  journal={Communications of the ACM},
  year={2020},
  volume={64},
  pages={58 - 65},
  url={https://api.semanticscholar.org/CorpusID:221655745}
}

@article{Furman2000TheDO,
  title={The Determinants of National Innovative Capacity},
  author={Jeffrey L. Furman and Michael E. Porter and Scott Stern},
  journal={IO: Productivity},
  year={2000},
  url={https://api.semanticscholar.org/CorpusID:3200579}
}

@article{Bloom2005IdentifyingTS,
  title={Identifying Technology Spillovers and Product Market Rivalry},
  author={Nick Bloom and Mark A. Schankerman and John Van Reenen},
  journal={Ewing Marion Kauffman Foundation Research Paper Series},
  year={2005},
  url={https://api.semanticscholar.org/CorpusID:241834}
}

@article{Porter2000MeasuringT,
  title={Measuring the "Ideas" Production Function: Evidence from International Patent Output},
  author={Michael Porter and Scott Stern},
  journal={IO: Productivity},
  year={2000},
  url={https://api.semanticscholar.org/CorpusID:154273158}
}

@article{Luintel2009HeterogeneousIP,
  title={Heterogeneous Ideas Production and Endogenous Growth: An Empirical Investigation},
  author={Kul B. Luintel and Mosahid Khan},
  journal={Wiley-Blackwell: Canadian Journal of Economics},
  year={2009},
  url={https://api.semanticscholar.org/CorpusID:23030763}
}

@article{Lanjouw2004PatentQA,
  title={Patent Quality and Research Productivity: Measuring Innovation with Multiple Indicators},
  author={Jean Olson Lanjouw and Mark A. Schankerman},
  journal={IO: Productivity},
  year={2004},
  url={https://api.semanticscholar.org/CorpusID:54495150}
}

\end{document}